\documentclass[11pt,a4paper]{article}

\usepackage{amsmath,amsthm,latexsym,amssymb}
\usepackage{graphicx,bm,fancybox,multirow,hhline,indentfirst,ulem}
\usepackage{ascmac,atbegshi}
\usepackage{natbib,verbatim,xcolor}
\usepackage[colorlinks,citecolor=blue,urlcolor=blue,dvipdfmx]{hyperref}

\makeatletter
\def\section{\@startsection {section}{1}{\z@}{-3.5ex plus -1ex minus -.2ex}{2.3 ex plus .2ex}{\normalsize\sc\centering}}
\def\subsection{\@startsection {subsection}{1}{\z@}{-3.5ex plus -1ex minus -.2ex}{2.3 ex plus .2ex}{\normalsize}}
\makeatother

\theoremstyle{definition}
\newtheorem{theorem}{Theorem}
\newtheorem*{lemma}{Lemma}

\newcommand{\indep}{\mathop{\perp\!\!\!\!\perp}}

\title{\vspace{-12mm}\normalsize\bf Doubly Robust Criterion for Causal Inference}
\author{\small Takamichi Baba\\[-1mm]\small Biostatistics Center, Shionogi \& Co., Ltd\\[-1mm]\small Department of Statistical Science, The Graduate University for Advanced Studies \and \small Yoshiyuki Ninomiya\thanks{\scriptsize Corresponding author. 10-3 Midori-cho, Tachikawa-shi, Tokyo 190-8562, Japan. E-mail: ninomiya@ism.ac.jp}\\[-1mm]\small Department of Statistical Inference and Mathematics, The Institute of Statistical Mathematics\\[-1mm]\small Department of Statistical Science, The Graduate University for Advanced Studies}
\date{}

\allowdisplaybreaks

\def\T{\prime}
\def\E{{\rm E}}
\def\P{{\rm P}}
\def\oP{{\rm o}_{\rm P}}
\def\OP{{\rm O}_{\rm P}}

\begin{document}

\renewcommand{\baselinestretch}{1.3}\selectfont

\maketitle

\vspace{-8mm}
\begin{abstract}
\noindent
The semiparametric estimation approach, which includes inverse-probability-weighted and doubly robust estimation using propensity scores, is a standard tool in causal inference, and it is rapidly being extended in various directions. On the other hand, although model selection is indispensable in statistical analysis, an information criterion for selecting an appropriate regression structure has just started to be developed. In this paper, based on the original definition of Akaike information criterion (AIC; \citealt{Aka73}), we derive an AIC-type criterion for propensity score analysis. Here, we define a risk function based on the Kullback-Leibler divergence as the cornerstone of the information criterion and treat a general causal inference model that is not necessarily a linear one. The causal effects to be estimated are those in the general population, such as the average treatment effect on the treated or the average treatment effect on the untreated. In light of the fact that this field attaches importance to doubly robust estimation, which allows either the model of the assignment variable or the model of the outcome variable to be wrong, we make the information criterion itself doubly robust so that either one can be wrong and it will still be an asymptotically unbiased estimator of the risk function. In simulation studies, we compare the derived criterion with an existing criterion obtained from a formal argument and confirm that the former outperforms the latter. Specifically, we check that the divergence between the estimated structure from the derived criterion and the true structure is clearly small in all simulation settings and that the probability of selecting the true or nearly true model is clearly higher. Real data analyses confirm that the results of variable selection using the two criteria differ significantly.

\medskip

\noindent 
Keywords: Generalized average treatment effect; Information criterion; Model selection; Propensity score analysis; Statistical asymptotic theory 
\end{abstract}



\section{Introduction}
\label{sec1}
Let us consider a fundamental setting in causal inference. It assumes that there are as many potential outcome variables as there are treatments, but that only the outcome variable corresponding to the assigned treatment is observed. There is confounding between the outcome and assignment, and if this is not taken into account and the estimation is done naively with the marginal likelihood, the estimates will have a bias. The bias can be avoided if the relationship between the outcome and confounding variables is modeled correctly, but a semiparametric approach without difficult modeling such as inverse-probability-weighted estimation (\citealt{RobRZ94}) or doubly robust estimation (\citealt{SchRR99}, \citealt{BanRob05}) is often taken.

For example, let $y_i^{(h)}\ (\in\{0,1\})$ be the potential outcome variable when the sample $i\ (\in\{1,2,\ldots,N\})$ is observed at time $h\ (\in\{1,2,\ldots,H\})$, $t_i^{(h)}$ be the assignment variable that becomes $1$ when $y_i^{(h)}$ is observed and $0$ when $y_i^{(h)}$ is not observed, $z_i\ (\in\mathbb{R})$ be the confounding variable. Furthermore, let us suppose a logistic model whose marginal probability function for the outcome variable is $f(y_i^{(h)};\theta)=p^{(h)y_i^{(h)}}(1-p^{(h)})^{1-y_i^{(h)}}$, where its regression structure is given by $\log\{p^{(h)}/(1-p^{(h)})\}=h\theta$, and whose conditional probability function for the assignment variable is $e^{(h)}(z_i)^{t_i^{(h)}}\{1-e^{(h)}(z_i)\}^{1-t_i^{(h)}}$ (see, for example, \citealt{HerR20}). Here, $e^{(h)}(z_i)$ is the propensity score introduced by \cite{RosRub83}, and $\theta$ is the parameter relating to the causal effect. In this model, if we obtain the estimator of $\theta$ by maximizing $\sum_{i=1}^Nt_i^{(h)}\log f(y_i^{(h)};\theta)$ despite the correlation between $y_i^{(h)}$ and $t_i^{(h)}$, it will have an asymptotic bias. The inverse-probability-weighted estimation is a method that uses the propensity scores to reproduce pseudo-complete data and gives an estimator by maximizing $\sum_{i=1}^Nt_i^{(h)}\log f(y_i^{(h)};\theta)/e^{(h)}(z_i)$.

In this statistical problem, it is reasonable to treat the model selection of the regression structure relating to the causal effect as constant or quadratic, rather than linear $h\theta$. However, for this basic selection, there are no reasonable information criteria. To be more specific, \cite{PlaBCWS13} pioneered the information criterion for such models, proposing the use of inverse-probability-weighting for the goodness-of-fit term. While it is quite appropriate, their criterion uses the number of parameters of the regression structure as a penalty term, which will lead to a considerable underestimation of the bias correction. \cite{BabKN17} asymptotically evaluated the bias and corrected the penalty term. However, what they proposed is a C$_p$-type criterion (\citealt{Mal73}) that can only handle linear models, such as $\E(y_i^{(h)})=h\theta$, and cannot deal with the problem described above. In addition, it can not deal with the average treatment effect on the treated (ATT) or the average treatment effect on the untreated (ATU), which is often dealt with in causal inference, i.e., the causal effect when the target population is not necessarily the whole. In this paper, based on the original definition of Akaike information criterion (AIC; \citealt{Aka73}), we derive an AIC-type criterion that overcomes these problems. If estimation using propensity scores is not used, the criterion proposed by \cite{RolY14} may be a reasonable alternative; however, it should be noted that the semiparametric approach is important and is what we have assumed here.

When the propensity score $e^{(h)}(z)$ is unknown, a simple method is to assume some parametric function $e^{(h)}(z;\alpha)$ as an approximation for it, obtain an estimator $\hat{\alpha}$ from $\{(t_i^{(h)},z_i):i\in\{1,2,\ldots,N\}\}$, and use $e^{(h)}(z;\hat{\alpha})$ instead of $e^{(h)}(z)$. On the other hand, if we suppose that the model may be misspecified, we can instead assume some parametric function $p^{(h)}(y^{(h)}\mid z;\beta)$ for the conditional distribution of the outcome variable given the confounding variable. Then, we can obtain an estimator $\hat{\beta}$ from $\{(t_i^{(h)},y_i^{(h)},z_i):i\in\{1,2,\ldots,N\}\}$, and use $p^{(h)}(y^{(h)}\mid z;\hat{\beta})$ to provide a doubly robust estimation; this is a standard method. If the modeling of either $e^{(h)}(z;\alpha)$ or $p^{(h)}(y^{(h)}\mid z;\beta)$ is correct, then $\theta$ can be estimated consistently. Although the information criterion given in \cite{BabKN17} covers doubly robust estimation, it is inadequate in that it is only derived when both modelings are correct. In this paper, we develop an information criterion that is an asymptotically unbiased estimator of an appropriate risk function as long as either modeling is correct, i.e., the information criterion itself is doubly robust. 

Standard extensions of AIC include TIC (\citealt{Tak76}) and GIC (\citealt{KonK96}); however, we would like to mention that our contribution does not fall within them. First, the penalty term in the AIC-type criterion is a certain asymptotic bias, so we usually have to consider convergence in mean, but to avoid unnecessary difficulties, we close the discussion with only weak convergence by using calculations peculiar to propensity score analysis. In the derivation of the doubly robust criterion, we use the asymptotic distribution of the doubly robust estimator under a setting that allows for model misspecification. To the best of our knowledge, there has been no effective use of that type of distribution. Moreover, the asymptotic bias depends both on the true distributions of the assignment and outcome variables, which may not be estimable in the setting of doubly robust estimation, and this is a difficulty not present in TIC and GIC; however, we propose a criterion that resolves this difficulty. It should also be emphasized that while the penalty terms of the traditional TIC and GIC are almost the same as twice the number of parameters, the penalty terms of our criterion tend to be much larger than that. This means that naively using the formal AIC can lead to the selection of significantly inappropriate models.

The organization of this paper is as follows. In Section \ref{sec2}, we describe a model, assumptions and a general causal effect in a setting where the target population is not necessarily the whole population, and introduce the inverse-probability-weighted estimator and the doubly robust estimator. In Section \ref{sec3}, we give a risk function based on the Kullback-Leibler divergence, which is naturally defined when considering such estimations, and define the AIC-type criterion by following the conventional derivation of AIC. Then, we asymptotically evaluate the penalty term of the information criterion for inverse-probability-weighted estimation. In Section \ref{sec4}, we derive the information criterion for doubly robust estimation, keeping in mind that one of the models of the assignment variable and the outcome variable may be misspecified. In Sections \ref{sec5} and \ref{sec6}, we compare the performance of the proposed criterion with the criterion devised by \cite{PlaBCWS13} through numerical experiments and an analysis based on real data. To explore the possibility of extending the derived information criterion, Section \ref{sec7} tries to generalize divergence and weight functions, in order to treat, for example, a loss function robust to outliers and a covariate balancing propensity score. Finally, Section \ref{sec8} summarizes our conclusions.

\section{Preparation}
\label{sec2}
\subsection{Model and assumption}
\label{sec2_1}
In this paper, we treat a fundamental causal inference model,
\begin{align*}
y = \sum_{h=1}^H t^{(h)} y^{(h)}, \qquad y^{(h)}\sim f(\cdot \mid \bm{x}^{(h)}; \bm{\theta}),
\end{align*}
where $t^{(h)}\ (\in\{0,1\})$ is an assignment variable that becomes $1$ when the $h$-th treatment is assigned ($\sum_{h=1}^H t^{(h)}=1$), $y^{(h)}\ (\in\mathbb{R})$ is a potential outcome variable when the $h$-th treatment is assigned, $\bm{x}^{(h)}\ (\in\mathbb{R}^r)$ is an explanatory variable for $y^{(h)}$, $f(y^{(h)} \mid \bm{x}^{(h)}; \bm{\theta})$ is the probability function of $y^{(h)}$ given $\bm{x}^{(h)}$, and $\bm{\theta}\ (\in \mathbb{R}^p)$ is the parameter used there ($h\in\{1,2,\ldots,H\}$). Note that $y$ on the left-hand side is an observed outcome variable. Also, $\bm{x}^{(h)}$ may contain some of the confounding variables, but for simplicity, the others are assumed to be non-random.

In this model, the $(H-1)$ potential outcome variables $y^{(h)}$'s with $t^{(h)}=0$ are regarded as missing. In addition, since $\E(y^{(h)})\neq \E(y^{(h)}\mid t^{(h)}=1)$ in general, naively estimating $\bm{\theta}$ from the observed values alone will result in a bias. Here, we suppose that the confounding variable $\bm{z}\ (\in\mathbb{R}^s)$ for $y^{(h)}$ and $t^{(h)}$ is observed such that this bias can be removed. Moreover, we assume a weakly ignorable treatment assignment condition,
\begin{align*}
y^{(h)} \indep t^{(h)} \mid \bm{z} \qquad (h\in\{1,2,\ldots,H\}),
\end{align*}
which is intended to allow for the removal of this bias (\citealt{Imb00}). We also assume positivity, $\P(t^{(h)}=1)>0$. There are $N$ samples following this model, and the variables in the $i$-th sample are denoted with the subscript $i$. Letting $\bm{u}_i=(y_i^{(1)},y_i^{(2)},\ldots,y_i^{(H)},t_i^{(1)},t_i^{(2)},\ldots,\allowbreak t_i^{(H)},\bm{x}_i^{(1)},\allowbreak\bm{x}_i^{(2)},\ldots,\bm{x}_i^{(H)},\bm{z}_i)$ be the variables, the samples are assumed to be independent, i.e.
\begin{align*}
\bm{u}_i \indep \bm{u}_j \qquad (i\neq j; \ i,j\in\{1,2,\ldots,N\}),
\end{align*}
which naturally implies that $y_{i}\indep y_{j}$.

The parameter of our interest to be estimated is $\bm{\theta}$ in a population where the $k$-th treatment group is $d^{(k)}$ times larger than it actually is, that is, it can be regarded as the parameter for estimating a generalized average treatment effect (\citealt{KalS22}). If $H=2$ and $(d^{(1)},d^{(2)})=(1,0)$ or $(d^{(1)},d^{(2)})=(0,1)$, $\bm{\theta}$ is a parameter of the average treatment effect on the treated or the untreated. If $H=2$ and $(d^{(1)},d^{(2)})=(1,1)$, $\bm{\theta}$ is a parameter of the average treatment effect of the whole sample. We denote the true value of the parameter by $\bm{\theta}^*$, and then it satisfies
\begin{align}
\sum_{h,k=1}^{H} \E \bigg\{ d^{(k)} t^{(k)} \frac{\partial}{\partial\bm{\theta}} \log f(y^{(h)} \mid \bm{x}^{(h)}; \bm{\theta}^*)\bigg\} = \bm{0}_p
\label{IPWEE}
\end{align}
where $\bm{0}_p$ is the $p$-dimensional zero vector. We may actually should write $(\partial/\partial \bm{\theta}) \log f_{\bm{\theta}^*}(y_i^{(h)} \mid \bm{x}_i^{(h)})$ as $(\partial/\partial \bm{\theta}) \log f_{\bm{\theta}}(y_i^{(h)} \mid \bm{x}_i^{(h)})|_{\bm{\theta}=\bm{\theta}^*}$, but for simplicity, this notation is used here and hereafter.

\subsection{Semiparametric estimation}
\label{sec2_2}
If the relationship between the potential outcome variable $y^{(h)}$ and the confounding variable $\bm{z}$ can be modeled correctly, then the causal effect can be estimated consistently from the maximum likelihood method under the ignorable treatment assignment condition; however, this modeling is generally difficult. In recent years, a semiparametric approach using the propensity score $\E(t^{(h)}\mid\bm{z})$, which does not necessarily require the above correct modeling, has often been used. Hereafter, a model for $\E(t^{(h)}\mid\bm{z})$ is denoted by $e^{(h)}(\bm{z};\bm{\alpha})$ with a further parameter $\bm{\alpha}\ (\in\mathbb{R}^q)$, and its true value is denoted by $\bm{\alpha}^*$. Below, we discuss two typical estimation methods that are used in this approach.

The first is inverse-probability-weighted estimation (\citealt{RobRZ94}). In this method, the missing values are pseudo-recovered by multiplying the observed values by the inverse of the propensity score as a weight, and then the usual estimation is implemented. Specifically, considering that the causal effect depends on $d^{(k)}$, we define a weighted loss function using the weights $w^{(h)}(\bm{z};\bm{\alpha}^*)\equiv\sum_{k=1}^Hd^{(k)}e^{(k)}(\bm{z};\bm{\alpha}^*)/e^{(h)}(\bm{z};\bm{\alpha}^*)$ and determine the inverse-probability-weighted estimator $\hat{\bm{\theta}}^{\rm IPW}$ by solving
\begin{align}
\frac{1}{N} \sum_{i=1}^N \sum_{h=1}^{H} t_i^{(h)} w^{(h)}(\bm{z}_i;\bm{\alpha}^*) \frac{\partial}{\partial\bm{\theta}} \log f(y_i^{(h)} \mid \bm{x}_i^{(h)}; \bm{\theta}) = \bm{0}_p.
\label{IPWEE2}
\end{align}
Conditional on $\bm{z}_i$, under the weakly ignorable treatment assignment condition, $t_i^{(h)}$ is independent of $y_i^{(h)}$ and has the expectation $e^{(h)}(\bm{z}_i;\bm{\alpha}^*)$, and then the left-hand side of \eqref{IPWEE2} converges to the left-hand side of \eqref{IPWEE}. This indicates that the inverse-probability-weighted estimator is consistent, i.e. $\hat{\bm{\theta}}^{\rm IPW}\stackrel{\rm p}{\to}\bm{\theta}^*$. Moreover, by substituting $\hat{\bm{\theta}}^{\rm IPW}$ into \eqref{IPWEE2} and expanding the result, we obtain
\begin{align}
\hat{\bm{\theta}}^{\rm IPW} - \bm{\theta}^* =
\frac{1}{N} \bm{A}(\bm{\theta}^*,\bm{\alpha}^*)^{-1} \sum_{i=1}^N \sum_{h=1}^H t_i^{(h)} w^{(h)}(\bm{z}_i;\bm{\alpha}^*) \frac{\partial}{\partial \bm{\theta}} \log f(y_i^{(h)} \mid \bm{x}_i^{(h)}; \bm{\theta}^*) \{1+\oP(1)\},
\label{IPWest}
\end{align}
where
\begin{align}
\bm{A}(\bm{\theta},\bm{\alpha}) \equiv \sum_{h,k=1}^{H} \E \bigg\{ - d^{(k)} e^{(k)}(\bm{z};\bm{\alpha}) \frac{\partial^2}{\partial\bm{\theta}\partial\bm{\theta}^{\T}} \log f(y^{(h)} \mid \bm{x}^{(h)}; \bm{\theta}) \bigg\}.
\label{Adef}
\end{align}
The above assumes that the propensity score is known, in other words, $\bm{\alpha}^*$ is known, but in general it is often unknown, in which case some estimator $\hat{\bm{\alpha}}$ is substituted for $\bm{\alpha}^*$. For $\hat{\bm{\alpha}}$, for example, we only have to construct the likelihood function for $t^{(h)}$ as a multinomial distribution with probability $e^{(h)}(\bm{z};\bm{\alpha})$ and use the maximum likelihood estimator.

Although $y^{(h)}$ is correlated with $\bm{z}$, the inverse-probability-weighted estimation does not directly use the information in $\bm{z}$ to estimate the expectation of $y^{(h)}$. The doubly robust estimation (\citealt{SchRR99}; \citealt{BanRob05}) improves the inverse-probability-weighted estimation by doing so. Denoting a model for the conditional distribution of $y^{(h)}$ given $\bm{z}$ by $p(y^{(h)} \mid \bm{z}; \bm{\beta})$ with a further parameter $\bm{\beta}\ (\in\mathbb{R}^r)$, and denoting its true value by $\bm{\beta}^*$, we take the expectation of $\log f(y^{(h)} \mid \bm{x}^{(h)}; \bm{\theta})$ with $p(y^{(h)}\mid \bm{z}; \bm{\beta})$ and write the conditional expectation as $g^{(h)}(\bm{x}^{(h)},\bm{z};\bm{\theta},\bm{\beta})$. In practice, some consistent estimator $\hat{\bm{\beta}}$ is usually substituted for $\bm{\beta}^*$. Specifically, if the propensity scores are unknown, we add
\begin{align}
\frac{1}{N} \sum_{i=1}^N \sum_{h=1}^{H} \bigg\{ \sum_{k=1}^{H}d^{(k)}t_i^{(k)}-t_i^{(h)}w^{(h)}(\bm{z}_i; \bm{\alpha}) \bigg\} \dfrac{\partial}{\partial\bm{\theta}} g^{(h)}(\bm{x}_i^{(h)},\bm{z}_i;\bm{\theta},\bm{\beta})
\label{DRadd}
\end{align}
to the left-hand side of (\ref{IPWEE2}), substitute estimators for $\bm{\alpha}$ and $\bm{\beta}$, and then find the doubly robust estimator $\hat{\bm{\theta}}^{\rm DR}$ by solving with respect to $\bm{\theta}$ such that it equals $\bm{0}_p$. This estimator not only improves the inverse-probability-weighted estimator but also is semiparametric locally efficient (\citealt{RobRot95}). In addition, if either the propensity score or the conditional expectation is specified correctly, it is consistent and hence said to be doubly robust.

\section{Inverse-probability-weighted criterion}
\label{sec3}
\subsection{Risk function for causal inference}
\label{sec3_1}
Before defining the risk function to derive the information criterion for causal inference, we will explain QIC$_w$ as proposed by \cite{PlaBCWS13}. QIC$_w$ is a criterion for missing data such as potential outcome variables, where the first term is the negative of twice the loss function used in the inverse-probability-weighted estimation, and the second term is twice the number of parameters. In the setting of Section \ref{sec2_2}, when $\bm{\alpha}^*$ is unknown, it is defined as
\begin{align}
{\rm QIC}_w = -2\sum_{i=1}^N \sum_{h=1}^H t_i^{(h)} w^{(h)}(\bm{z}_i;\hat{\bm{\alpha}}) \log f(y_i^{(h)} \mid \bm{x}_i^{(h)}; \hat{\bm{\theta}}^{\rm IPW}) + 2p.
\label{qicw}
\end{align}
When there are no weights $t_i^{(h)} w^{(h)}(\bm{z}_i;\hat{\bm{\alpha}})$ in QIC$_w$, QIC$_w$ is an asymptotic unbiased estimator of the risk function based on the Kullback-Leibler divergence. On the other hand, when weights are present, the variance of the first term increases, and the second term should be made larger.

Similarly to the usual AIC-type criteria, let $(\tilde{y}_i^{(h)},\tilde{t}_i^{(h)},\tilde{\bm{x}}_i^{(h)},\tilde{\bm{z}}_i)$ be a copy of $(y_i^{(h)},t_i^{(h)},\bm{x}_i^{(h)},\bm{z}_i)$, i.e., a random vector that is independently and identically distributed according to the distribution of $(y_i^{(h)},t_i^{(h)},\bm{x}_i^{(h)},\bm{z}_i)$. Furthermore, let $\hat{\bm{\theta}}$ be an appropriate consistent estimator of $\bm{\theta}$ with square root of $n$ convergence, where we assume that we can write $n^{1/2}(\hat{\bm{\theta}}-\bm{\theta}^*)=\bm{\xi}+\oP(1)$ using a random vector $\bm{\xi}$ which is $\OP(1)$ and whose expectation is $0$. Here, we can define a risk function
\begin{align}
-2 \sum_{i=1}^N \sum_{h=1}^H \E \{ \tilde{t}_i^{(h)} w^{(h)}(\tilde{\bm{z}}_i) \log f(\tilde{y}_i^{(h)} \mid \tilde{\bm{x}}_i^{(h)}; \hat{\bm{\theta}}) \}
\label{risk}
\end{align}
such that the first term of \eqref{qicw} is naturally used as its naive estimator, where $w^{(h)}(\bm{z})$ is the limit of $w^{(h)}(\bm{z};\hat{\bm{\alpha}})$. Although some may be willing to use its true value, $\sum_{k=1}^Hd^{(k)}\E(t^{(k)}\mid\bm{z})/\E(t^{(h)}\mid\bm{z})$, instead of the limit, here we consider the risk more matching to $\hat{\bm{\alpha}}$, which we are actually using. As we will see later, in fact, using the true value cannot provide a practical information criterion for doubly robust estimation. Note that when $\bm{\alpha}^*$ is known, we use $w^{(h)}(\bm{z};\bm{\alpha}^*)$ as $w^{(h)}(\bm{z})$. This risk function can be regarded to be based on the Kullback-Leibler divergence between the true and estimated distributions in a population where the $k$-th treatment group is $d^{(k)}$ times larger than it actually is. Also, as can be seen from \eqref{IPWEE2}, the estimation aims to minimize this risk function, and in that sense, it is a natural quantity. Since the first term in \eqref{qicw} is an evaluation of this risk function using the same data as those for the estimator, it tends to be smaller than the true value. Therefore, denoting the bias as
\begin{align}
& -2 \E \bigg[ \sum_{i=1}^N \sum_{h=1}^H t_i^{(h)} w^{(h)}(\bm{z}_i) \log f(y_i^{(h)} \mid \bm{x}_i^{(h)}; \hat{\bm{\theta}}) 
- \sum_{i=1}^N \sum_{h=1}^H \E \{ \tilde{t}_i^{(h)} w^{(h)}(\tilde{\bm{z}}_i) \log f(\tilde{y}_i^{(h)} \mid \tilde{\bm{x}}_i^{(h)}; \hat{\bm{\theta}}) \} \bigg]
\notag \\
& = -2 \E \bigg[ \sum_{i=1}^N \sum_{h=1}^H t_i^{(h)} w^{(h)}(\bm{z}_i) \{ \log f(y_i^{(h)} \mid \bm{x}_i^{(h)}; \hat{\bm{\theta}}) - \log f(y_i^{(h)} \mid \bm{x}_i^{(h)}; \bm{\theta}^*) \}
\notag \\
& \phantom{= -2 \E \bigg[} - \sum_{i=1}^N \sum_{h=1}^H \tilde{t}_i^{(h)} w^{(h)}(\tilde{\bm{z}}_i) \{ \log f(\tilde{y}_i^{(h)} \mid \tilde{\bm{x}}_i^{(h)}; \hat{\bm{\theta}}) - \log f(\tilde{y}_i^{(h)} \mid \tilde{\bm{x}}_i^{(h)}; \bm{\theta}^*) \} \bigg],
\label{bias}
\end{align}
we define $b^{\rm limit}$ as the weak limit of the quantity in the expectation in \eqref{bias}, i.e., the random variable to which the quantity converges in distribution, and use $\E(b^{\rm limit})$ as the asymptotic bias for the correction. If \eqref{bias} is expanded with respect to $\hat{\bm{\theta}}$ around $\bm{\theta}^*$, we get
\begin{align*}
& b^{\rm limit} = -2 \bm{\xi}^{\T} \sum_{i=1}^N \sum_{h=1}^H t_i^{(h)} w^{(h)}(\bm{z}_i) \dfrac{\partial}{\partial \bm{\theta}} \log f(y_i^{(h)} \mid \bm{x}_i^{(h)}; \bm{\theta}^*)
\\
& \phantom{b^{\rm limit} =} +2 \bm{\xi}^{\T} \sum_{i=1}^N \sum_{h=1}^H \tilde{t}_i^{(h)} w^{(h)}(\tilde{\bm{z}}_i) \dfrac{\partial}{\partial \bm{\theta}} \log f(\tilde{y}_i^{(h)} \mid \tilde{\bm{x}}_i^{(h)}; \bm{\theta}^*).
\end{align*}
The expectation of the second term is divided into the expectation of $\bm{\xi}$ and the expectation of the other terms; the former is $\bm{0}_p$, so we obtain
\begin{align}
& \E (b^{\rm limit}) = -2 \sum_{i=1}^N \sum_{h=1}^H \E \bigg\{ \bm{\xi}^{\T} t_i^{(h)} w^{(h)}(\bm{z}_i) \dfrac{\partial}{\partial \bm{\theta}} \log f(y_i^{(h)} \mid \bm{x}_i^{(h)}; \bm{\theta}^*) \bigg\}.
\label{bias2}
\end{align}
When this asymptotic bias contains unknown parameters, they are replaced by their consistent estimators, as is done in deriving the usual AIC-type information criterion, and the evaluated asymptotic bias is added to the first term of \eqref{qicw} to construct the information criterion.

This construction of the information criterion is based on \cite{BabKN17}, but the same kind of construction is used in \cite{WalMS19}, so we would like to mention the differences between the two. \cite{WalMS19} proposes a valuable information criterion that is also justified in causal inference, and actually the main focus is on its application to g-estimation. On the other hand, the derivation is intuitive, and the information criterion is defined by substituting an estimator into the expectation of quasi-log-likelihood. Specifically, letting $Q(\cdot)$ be the quasi-log-likelihood, $\hat{\bm{\psi}}$ be the estimator, and $\bm{\psi}^{\dagger}$ be the limit of the estimator, then $-2Q(\hat{\bm{\psi}})$ is used instead of $\E\{-2Q(\bm{\psi}^{\dagger})\}$ without considering the bias. In this paper, while being conscious of the difference between convergence in mean and weak convergence, we use calculations peculiar to propensity score analysis based on the ignorable treatment assignment condition, and construct the information criterion in a way that may seem somewhat complicated. As will become clear later, while the penalty term of the criterion in \cite{WalMS19} for g-estimation should be close to that of AIC, the penalty term of our criterion for estimation based on inverse-probability-weighting tends to be much larger than that of AIC.

\subsection{Case of known propensity score}
\label{sec3_2}
Since the error of the inverse-probability-weighted estimator can be written as \eqref{IPWest}, we hereafter evaluate
\begin{align}
\E (b^{\rm limit}) = -\frac{2}{N} \sum_{i,j=1}^N \sum_{h,k=1}^H \E \bigg[ & \bigg\{ t_i^{(h)} w^{(h)}(\bm{z}_i;\bm{\alpha}^*) \frac{\partial}{\partial \bm{\theta}^{\T}} \log f(y_i^{(h)}\mid\bm{x}_i^{(h)};\bm{\theta}^*) \bigg\} 
\notag \\
& \bm{A}(\bm{\theta}^*,\bm{\alpha}^*)^{-1} \bigg\{ t_j^{(k)} w^{(k)}(\bm{z}_j;\bm{\alpha}^*) \frac{\partial}{\partial \bm{\theta}} \log f(y_j^{(k)}\mid\bm{x}_j^{(k)};\bm{\theta}^*) \bigg\} \bigg].
\label{blimit1}
\end{align}
When $i\neq j$, this expectation is divided into one for $i$ and one for $j$ because of their independence, and the former is summed over $h$ to get
\begin{align}
& \E \bigg\{ \sum_{h=1}^H t_i^{(h)} w^{(h)}(\bm{z}_i;\bm{\alpha}^*) \frac{\partial}{\partial \bm{\theta}} \log f(y_i^{(h)} \mid \bm{x}_i^{(h)}; \bm{\theta}^*) \bigg\}
\notag \\
& = \E_{\bm{z}_i} \bigg[ \sum_{h=1}^H \E \{ t_i^{(h)} w^{(h)}(\bm{z}_i;\bm{\alpha}^*) \mid \bm{z}_i \} \E \bigg\{ \frac{\partial}{\partial \bm{\theta}} \log f(y_i^{(h)} \mid \bm{x}_i^{(h)}; \bm{\theta}^*) \ \bigg| \ \bm{z}_i \bigg\} \bigg]
\notag \\
& = \E_{\bm{z}_i} \bigg[ \sum_{h=1}^H \E \bigg\{ \sum_{k=1}^H d^{(k)} e^{(k)}(\bm{z}_i;\bm{\alpha}^*) \bigg\} \E \bigg\{ \frac{\partial}{\partial \bm{\theta}} \log f(y_i^{(h)} \mid \bm{x}_i^{(h)}; \bm{\theta}^*) \ \bigg| \ \bm{z}_i \bigg\} \bigg]
\notag \\
& = \E \bigg\{ \sum_{k,h=1}^H d^{(k)} e^{(k)}(\bm{z}_i;\bm{\alpha}^*) \frac{\partial}{\partial \bm{\theta}} \log f(y_i^{(h)} \mid \bm{x}_i^{(h)}; \bm{\theta}^*) \bigg\}
\notag \\
& = \bm{0}_p.
\label{base1}
\end{align}
The last equality is derived from \eqref{IPWEE}. Therefore, the expectation in the right-hand side of \eqref{blimit1} becomes the trace of 
\begin{align*}
\bm{A}(\bm{\theta}^*,\bm{\alpha}^*)^{-1} \E \bigg\{ t_i^{(h)} w^{(h)}(\bm{z}_i;\bm{\alpha}^*) t_i^{(k)} w^{(k)}(\bm{z}_i;\bm{\alpha}^*) \frac{\partial}{\partial \bm{\theta}} \log f(y_i^{(h)}\mid\bm{x}_i^{(h)};\bm{\theta}^*) \frac{\partial}{\partial \bm{\theta}^{\T}} \log f(y_i^{(k)}\mid\bm{x}_i^{(k)};\bm{\theta}^*) \bigg\}.
\end{align*}
When $k\neq h$, the components are $0$ because $t_i^{(h)} t_i^{(k)}=0$. When $k=h$, this expectation becomes
\begin{align}
& \E \bigg\{ t_i^{(h)} w^{(h)}(\bm{z}_i;\bm{\alpha}^*)^2 \frac{\partial}{\partial \bm{\theta}} \log f(y_i^{(h)} \mid \bm{x}_i^{(h)}; \bm{\theta}^*) \frac{\partial}{\partial \bm{\theta}^{\T}} \log f(y_i^{(h)} \mid \bm{x}_i^{(h)}; \bm{\theta}^*) \bigg\}
\notag \\
& = \E_{\bm{z}_i} \bigg[ \E ( t_i^{(h)} \mid \bm{z}_i ) w^{(h)}(\bm{z}_i;\bm{\alpha}^*)^2 \E \bigg\{ \frac{\partial}{\partial \bm{\theta}} \log f(y_i^{(h)} \mid \bm{x}_i^{(h)}; \bm{\theta}^*) \frac{\partial}{\partial \bm{\theta}^{\T}} \log f(y_i^{(h)} \mid \bm{x}_i^{(h)}; \bm{\theta}^*) \ \bigg| \ \bm{z}_i \bigg\} \bigg]
\notag \\
& = \E \bigg[ \bigg\{\sum_{k=1}^Hd^{(k)}e^{(k)}(\bm{z}_i;\bm{\alpha}^*)\bigg\}^2 \frac{1}{e^{(h)}(\bm{z}_i;\bm{\alpha}^*)} \frac{\partial}{\partial \bm{\theta}} \log f(y_i^{(h)} \mid \bm{x}_i^{(h)}; \bm{\theta}^*) \frac{\partial}{\partial \bm{\theta}^{\T}} \log f(y_i^{(h)} \mid \bm{x}_i^{(h)}; \bm{\theta}^*) \bigg].
\label{pre1}
\end{align}
Therefore, by defining 
\begin{align}
\bm{B}(\bm{\theta},\bm{\alpha}) \equiv \sum_{h=1}^H \E \bigg[ \bigg\{\sum_{k=1}^Hd^{(k)}e^{(k)}(\bm{z};\bm{\alpha})\bigg\}^2 \frac{1}{e^{(h)}(\bm{z};\bm{\alpha})} \frac{\partial}{\partial \bm{\theta}} \log f(y^{(h)} \mid \bm{x}^{(h)}; \bm{\theta}) \frac{\partial}{\partial \bm{\theta}^{\T}} \log f(y^{(h)} \mid \bm{x}^{(h)}; \bm{\theta}) \bigg],
\label{Bdef}
\end{align}
we arrive at the following theorem.

\begin{theorem}
Suppose that $\bm{A}(\bm{\theta},\bm{\alpha})$ and $\bm{B}(\bm{\theta},\bm{\alpha})$ are defined as in \eqref{Adef} and \eqref{Bdef}. Then, the asymptotic bias of the information criterion for the inverse-probability-weighted estimation is given by
\begin{align*}
\E (b^{\rm limit}) = -2 \mathrm{tr} \{ \bm{A}(\bm{\theta}^*,\bm{\alpha}^*)^{-1} \bm{B}(\bm{\theta}^*,\bm{\alpha}^*) \}
\end{align*}
when the propensity score is known.
\label{th1}
\end{theorem}

On the basis of this result and the fact that $\hat{\bm{\theta}}^{\rm IPW}$ is a consistent estimator of $\bm{\theta}$, we propose
\begin{align*}
\mathrm{IPWIC} \equiv -2 \sum_{i=1}^N \sum_{h=1}^{H} t_i^{(h)} w^{(h)}(\bm{z}_i;\bm{\alpha}^*) \log f(y_i^{(h)} \mid \bm{x}_i^{(h)}; \hat{\bm{\theta}}^{\mathrm{IPW}}) + 2 \mathrm{tr} \{ \hat{\bm{A}}(\hat{\bm{\theta}}^{\mathrm{IPW}},\bm{\alpha}^*)^{-1} \hat{\bm{B}}(\hat{\bm{\theta}}^{\mathrm{IPW}},\bm{\alpha}^*) \}
\end{align*}
as an information criterion for the inverse-probability-weighted estimation when the propensity score is known, where $\hat{\bm{A}}(\bm{\theta},\bm{\alpha})$ and $\hat{\bm{B}}(\bm{\theta},\bm{\alpha})$ are empirical version of $\bm{A}(\bm{\theta},\bm{\alpha})$ and $\bm{B}(\bm{\theta},\bm{\alpha})$, respectively. Note that if $d^{(1)}=d^{(2)}=\cdots=d^{(H)}=1$, the matrix made by removing $1/e^{(h)}(\bm{z};\bm{\alpha})$ from the definition of $\hat{\bm{B}}(\bm{\theta},\bm{\alpha})$ is the same as $\hat{\bm{A}}(\bm{\theta},\bm{\alpha})$. It means that if our target of estimation is close to the average treatment effect of the whole sample, the IPWIC penalty term will be much larger than twice the number of parameters.

\subsection{Case of unknown propensity score}
\label{sec3_3}
If the parameter $\bm{\alpha}$ in the propensity score $e^{(h)}(\bm{z};\bm{\alpha})$ is unknown, then we only have to maximize the log-likelihood $\sum_{i=1}^{N} \sum_{h=1}^{H} t_i^{(h)} \log e^{(h)}(\bm{z}_i; \bm{\alpha})$ to find $\hat{\bm{\alpha}}$. From this log-likelihood, the score function is $\sum_{i=1}^{N}\sum_{h=1}^{H}\{ t_i^{(h)} / e^{(h)}(\bm{z}_i; \bm{\alpha}^*)\} \{\partial e^{(h)}(\bm{z}_i; \bm{\alpha}^*) / \partial \bm{\alpha}\}$, the Fisher information matrix is
\begin{align}
\bm{I}_1(\bm{\alpha}^*) \equiv \sum_{h=1}^H \E \bigg\{ \frac{1}{e^{(h)}(\bm{z}; \bm{\alpha}^*)} \frac{\partial}{\partial \bm{\alpha}} e^{(h)}(\bm{z}; \bm{\alpha}^*) \frac{\partial}{\partial \bm{\alpha}^{\T}} e^{(h)}(\bm{z}; \bm{\alpha}^*) \bigg\},
\label{defI1}
\end{align}
and the error of $\hat{\bm{\alpha}}$ is expressed as
\begin{align*}
\hat{\bm{\alpha}} - \bm{\alpha}^* = \frac{1}{N} \bm{I}_1(\bm{\alpha}^*)^{-1} \sum_{h=1}^{H} \sum_{i=1}^{N} t_i^{(h)} \frac{\partial}{\partial \bm{\alpha}} \log e^{(h)}(\bm{z}_i; \bm{\alpha}^*) \{1+\oP(1)\}.
\end{align*}
From this and \eqref{IPWest}, the error of the inverse-probability-weighted estimator is
\begin{align}
& \hat{\bm{\theta}}^{\rm IPW} - \bm{\theta}^*
\notag \\
& = \frac{1}{N} \bm{A}(\bm{\theta}^*,\hat{\bm{\alpha}})^{-1}
\notag \\
& \phantom{=} \ \sum_{i=1}^N \sum_{h=1}^H t_i^{(h)} \bigg\{ w^{(h)}(\bm{z}_i; \bm{\alpha}^*) + \frac{\partial}{\partial \bm{\alpha}^{\T}} w^{(h)}(\bm{z}_i; \bm{\alpha}^*) (\hat{\bm{\alpha}}-\bm{\alpha}^*) \bigg\} \frac{\partial}{\partial \bm{\theta}} \log f(y_i^{(h)} \mid \bm{x}_i^{(h)}; \bm{\theta}^*) \{1+\oP(1)\}
\notag \\
& = \frac{1}{N} \bm{A}(\bm{\theta}^*,\bm{\alpha}^*)^{-1} \sum_{i=1}^N \sum_{h=1}^H \bigg\{ t_i^{(h)} w^{(h)}(\bm{z}_i; \bm{\alpha}^*) \frac{\partial}{\partial \bm{\theta}} \log f(y_i^{(h)} \mid \bm{x}_i^{(h)}; \bm{\theta}^*)
\notag \\
& \phantom{= \bm{A}(\bm{\theta}^*,\bm{\alpha})^{-1} \frac{1}{N} \sum_{i=1}^N \sum_{h=1}^H \bigg\{} - \bm{\Lambda}_1(\bm{\theta}^*,\bm{\alpha}^*)^{\T} \bm{I}_1(\bm{\alpha}^*)^{-1} t_i^{(h)} \frac{\partial}{\partial \bm{\alpha}} \log e^{(h)}(\bm{z}_i; \bm{\alpha}^*) \bigg\} \{1+\oP(1)\},
\label{IPWest2}
\end{align}
where
\begin{align}
\bm{\Lambda}_1(\bm{\theta},\bm{\alpha}) \equiv \sum_{h=1}^H \E \bigg\{ -e^{(h)}(\bm{z}; \bm{\alpha}) \frac{\partial}{\partial \bm{\alpha}} w^{(h)}(\bm{z}; \bm{\alpha}) \frac{\partial}{\partial \bm{\theta}^{\T}} \log f(y^{(h)} \mid \bm{x}^{(h)}; \bm{\theta}) \bigg\}.
\label{defLam1}
\end{align}
Since the error can be written like this, on the basis of \eqref{bias2}, we hereafter evaluate
\begin{align*}
& \E (b^{\rm limit})
\\
& = -\frac{2}{N} \sum_{i,j=1}^N \sum_{h,k=1}^H {\rm tr} \bigg( \bm{A}(\bm{\theta}^*,\bm{\alpha}^*)^{-1} \E \bigg[ t_i^{(h)} w^{(h)}(\bm{z}_i; \bm{\alpha}^*) \frac{\partial}{\partial \bm{\theta}} \log f(y_i^{(h)}\mid\bm{x}_i^{(h)};\bm{\theta}^*)
\\
& \phantom{=} \bigg\{ t_j^{(k)} w^{(k)}(\bm{z}_j; \bm{\alpha}^*) \frac{\partial}{\partial \bm{\theta}^{\T}} \log f(y_j^{(k)} \mid \bm{x}_j^{(k)}; \bm{\theta}^*) - t_j^{(k)} \frac{\partial}{\partial \bm{\alpha}^{\T}} \log e^{(k)}(\bm{z}_j; \bm{\alpha}^*) \bm{I}_1(\bm{\alpha}^*)^{-1} \bm{\Lambda}_1(\bm{\theta}^*,\bm{\alpha}^*) \bigg\} \bigg] \bigg).
\end{align*}
When $i\neq j$, this expectation is divided into one for $i$ and one for $j$ because of their independence, and the sum of the expectations of the terms with $i$ taken over $h$ is $\bm{0}_p$ from \eqref{base1}. Therefore, we only have to evaluate the terms with $i=j$. When $k\neq h$, the components are $0$ because $t_i^{(h)} t_i^{(k)}= 0$. On the other hand, when $k=h$, the first term is the same as \eqref{pre1}. The next term is
\begin{align*}
& \E \bigg\{ t_i^{(h)} w^{(h)}(\bm{z}_i; \bm{\alpha}^*) \frac{\partial}{\partial \bm{\theta}} \log f(y_i^{(h)} \mid \bm{x}_i^{(h)}; \bm{\theta}^*) \frac{\partial}{\partial \bm{\alpha}^{\T}} \log e^{(h)}(\bm{z}_i; \bm{\alpha}^*) \bm{I}_1(\bm{\alpha}^*)^{-1} \bm{\Lambda}_1(\bm{\theta}^*,\bm{\alpha}^*) \bigg\}
\\
& = \E \bigg\{ w^{(h)}(\bm{z}_i; \bm{\alpha}^*) \frac{\partial}{\partial \bm{\theta}} \log f(y_i^{(h)} \mid \bm{x}_i^{(h)}; \bm{\theta}^*) \frac{\partial}{\partial \bm{\alpha}^{\T}} e^{(h)}(\bm{z}_i; \bm{\alpha}^*) \bm{I}_1(\bm{\alpha}^*)^{-1} \bm{\Lambda}_1(\bm{\theta}^*,\bm{\alpha}^*) \bigg\}.
\end{align*}
By defining
\begin{align}
\bm{\Lambda}_2(\bm{\theta},\bm{\alpha}) \equiv \sum_{h=1}^H \E \bigg\{ w^{(h)}(\bm{z}; \bm{\alpha}) \frac{\partial}{\partial \bm{\theta}} \log f(y^{(h)} \mid \bm{x}^{(h)}; \bm{\theta}) \frac{\partial}{\partial \bm{\alpha}^{\T}} e^{(h)}(\bm{z}; \bm{\alpha}) \bigg\},
\label{defLam2}
\end{align}
we arrive at the following theorem.

\begin{theorem}
Suppose that $\bm{A}(\bm{\theta},\bm{\alpha})$, $\bm{B}(\bm{\theta},\bm{\alpha})$, $\bm{I}_1(\bm{\alpha})$, $\bm{\Lambda}_1(\bm{\theta},\bm{\alpha})$ and $\bm{\Lambda}_2(\bm{\theta},\bm{\alpha})$ are defined as in \eqref{Adef}, \eqref{Bdef}, \eqref{defI1}, \eqref{defLam1} and \eqref{defLam2}. Then, the asymptotic bias of the information criterion for the inverse-probability-weighted estimation is given by
\begin{align*}
\E (b^{\rm limit}) = -2 \mathrm{tr} \{ \bm{A}(\bm{\theta}^*,\bm{\alpha}^*)^{-1} \bm{B}(\bm{\theta}^*,\bm{\alpha}^*) - \bm{\Lambda}_2(\bm{\theta}^*,\bm{\alpha}^*) \bm{I}_1(\bm{\alpha}^*)^{-1} \bm{\Lambda}_1(\bm{\theta}^*,\bm{\alpha}^*) \}
\end{align*}
when the propensity score is unknown.
\label{th2}
\end{theorem}

On the basis of this result and the fact that $(\hat{\bm{\theta}}^{\mathrm{IPW}},\hat{\bm{\alpha}})$ is a consistent estimator of $(\bm{\theta}^*,\bm{\alpha}^*)$, we propose
\begin{align}
& \mathrm{IPWIC} \equiv -2 \sum_{i=1}^N \sum_{h=1}^{H} t_i^{(h)} w^{(h)}(\bm{z}_i,\hat{\bm{\alpha}}) \log f(y_i^{(h)} \mid \bm{x}_i^{(h)}; \hat{\bm{\theta}}^{\mathrm{IPW}})
\notag \\
& \phantom{\mathrm{IPWIC} \equiv} + 2 \mathrm{tr} \{ \hat{\bm{A}}(\hat{\bm{\theta}}^{\mathrm{IPW}},\hat{\bm{\alpha}})^{-1} \hat{\bm{B}}(\hat{\bm{\theta}}^{\mathrm{IPW}},\hat{\bm{\alpha}}) - \hat{\bm{\Lambda}}_2(\hat{\bm{\theta}}^{\mathrm{IPW}},\hat{\bm{\alpha}}) \hat{\bm{I}}_1(\hat{\bm{\alpha}})^{-1} \hat{\bm{\Lambda}}_1(\hat{\bm{\theta}}^{\mathrm{IPW}},\hat{\bm{\alpha}})\}
\label{ipwic2}
\end{align}
as an information criterion for the inverse-probability-weighted estimation when the propensity score is unknown, where $\hat{\bm{I}}_{1}(\bm{\alpha})$, $\hat{\bm{\Lambda}}_{1}(\bm{\theta},\bm{\alpha})$ and $\hat{\bm{\Lambda}}_{2}(\bm{\theta},\bm{\alpha})$ are empirical version of $\bm{I}_{1}(\bm{\alpha})$, $\bm{\Lambda}_{1}(\bm{\theta},\bm{\alpha})$ and $\bm{\Lambda}_{2}(\bm{\theta},\bm{\alpha})$, respectively. Theorems \ref{th1} and \ref{th2} indicate that the penalty for an unknown propensity score tends to be smaller than that for a known propensity score. This is consistent with the fact that the asymptotic variance for the inverse-probability-weighted estimator becomes smaller if the propensity score is estimated even when it is known (see, e.g., \citealt{HenEgu04}).

\section{Doubly robust criterion}
\label{sec4}
The doubly robust estimator is an estimator that is consistent even if either the model for the assignment variable conditional on the confounding variable or the model for the outcome variable conditional on the confounding variable is misspecified. In the narrow setting of a linear model and a basic average treatment effect, \cite{BabKN17} proposed a C$_p$-type criterion for the doubly robust estimation, but derived it under the assumption that both models are correct, so it has no validity when only one of them is correct. In this section, we aim to derive a doubly robust criterion that is an asymptotically unbiased estimator of the risk function in \eqref{risk}. Note that in the setting of doubly robust estimation, the propensity score is unknown; i.e., $\bm{\alpha}^*$ is unknown.

Letting $\bm{u}=(y^{(1)},y^{(2)},\ldots,y^{(H)},t^{(1)},t^{(2)},\ldots,t^{(H)},\bm{x}^{(1)},\bm{x}^{(2)},\ldots,\bm{x}^{(H)},\bm{z})$, we define
\begin{align}
\bm{m} (\bm{u};\bm{\theta},\bm{\alpha},\bm{\beta}) \equiv & \sum_{h=1}^H \frac{\partial}{\partial \bm{\theta}} \bigg[ t^{(h)} w^{(h)}(\bm{z};\bm{\alpha}) \log f(y^{(h)} \mid \bm{x}^{(h)}; \bm{\theta}) 
\notag\\
& \phantom{\sum_{h=1}^H \frac{\partial}{\partial \bm{\theta}} \bigg[} + \bigg\{ \sum_{k=1}^{H}d^{(k)} t^{(k)} - t^{(h)} w^{(h)}(\bm{z}; \bm{\alpha}) \bigg\} g^{(h)}(\bm{x}^{(h)},\bm{z};\bm{\theta},\bm{\beta}) \bigg].
\label{mdef}
\end{align}
Then, the doubly robust estimating equation is 
\begin{align*}
\sum_{i=1}^N 
\begin{pmatrix}
\bm{m} (\bm{u}_i;\bm{\theta},\bm{\alpha},\bm{\beta})
\\
\displaystyle \sum_{h=1}^H \frac{\partial}{\partial \bm{\alpha}} t_i^{(h)} \log e^{(h)}(\bm{z}_i; \bm{\alpha}) 
\\
\displaystyle \sum_{h=1}^H \frac{\partial}{\partial \bm{\beta}} t_i^{(h)} \log p^{(h)} (\bm{y}_i^{(h)} \mid \bm{z}_i; \bm{\beta})
\end{pmatrix}
= \bm{0}_{p+q+r}.
\end{align*}
Let us denote the limit of the solution $(\hat{\bm{\theta}}^{\rm DR}, \hat{\bm{\alpha}}, \hat{\bm{\beta}})$ of this estimating equation by $(\bm{\theta}^{\dagger}, \bm{\alpha}^{\dagger}, \bm{\beta}^{\dagger})$. If one of the models is correct, $\bm{\theta}^{\dagger}=\bm{\theta}^*$ (we will write $\bm{\theta}^*$ from now on), but not necessarily $\bm{\alpha}^{\dagger}=\bm{\alpha}^*$ or $\bm{\beta}^{\dagger}=\bm{\beta}^*$. We perform a Taylor expansion on the estimating equation with the estimator substituted into it and develop a statistical asymptotic theory similar to the conventional one. The derivative of the left-hand side of the estimating equation is asymptotically the sum of 
\begin{align*} 
\begin{pmatrix}
\dfrac{\partial}{\partial \bm{\theta}^{\T}} \bm{m} (\bm{u}_i;\bm{\theta}^*,\bm{\alpha}^{\dagger},\bm{\beta}^{\dagger}) & \dfrac{\partial}{\partial \bm{\alpha}^{\T}} \bm{m} (\bm{u}_i;\bm{\theta}^*,\bm{\alpha}^{\dagger},\bm{\beta}^{\dagger}) & \dfrac{\partial}{\partial \bm{\beta}^{\T}} \bm{m} (\bm{u}_i;\bm{\theta}^*,\bm{\alpha}^{\dagger},\bm{\beta}^{\dagger}) \\
\bm{O} & \displaystyle \sum_{h=1}^H \dfrac{\partial^2}{\partial \bm{\alpha} \partial \bm{\alpha}^{\T}} t_i^{(h)} \log e^{(h)}(\bm{z}_i; \bm{\alpha}^{\dagger}) & \bm{O} \\
\bm{O} & \bm{O} & \displaystyle \sum_{h=1}^H \dfrac{\partial^2}{\partial \bm{\beta} \partial \bm{\beta}^{\T}} t_i^{(h)} \log p^{(h)} (\bm{y}_i^{(h)} \mid \bm{z}_i; \bm{\beta}^{\dagger})
\end{pmatrix},
\end{align*}
and if we divide it by $-N$, it converges in probability to 
\begin{align} 
\begin{pmatrix}
\E \bigg\{ -\dfrac{\partial}{\partial \bm{\theta}^{\T}} \bm{m} (\bm{u}; \bm{\theta}^*,\bm{\alpha}^{\dagger},\bm{\beta}^{\dagger}) \bigg\} & \E \bigg\{ -\dfrac{\partial}{\partial \bm{\alpha}^{\T}} \bm{m} (\bm{u};\bm{\theta}^*,\bm{\alpha}^{\dagger},\bm{\beta}^{\dagger}) \bigg\} & \E \bigg\{ -\dfrac{\partial}{\partial \bm{\beta}^{\T}} \bm{m} (\bm{u};\bm{\theta}^*,\bm{\alpha}^{\dagger},\bm{\beta}^{\dagger}) \bigg\}
\\
\bm{O} & \bm{I}_1(\bm{\alpha}^{\dagger}) & \bm{O}
\\
\bm{O} & \bm{O} & \bm{I}_2(\bm{\beta}^{\dagger})
\end{pmatrix}
\label{mat1}
\end{align}
from the law of large numbers. Here, $\bm{I}_1(\cdot)$ and $\bm{I}_2(\cdot)$ are the Fisher information matrices for $\bm{\alpha}$ and $\bm{\beta}$, respectively. and $\bm{O}$ is a zero matrix. The terms in the (1,1) block are
\begin{align*}
& \sum_{h=1}^H \E_{\bm{z}} \bigg[ \E \{ t^{(h)} w^{(h)}(\bm{z};\bm{\alpha}^{\dagger}) \mid \bm{z} \} \E \bigg\{ -\dfrac{ \partial^2}{\partial \bm{\theta} \partial \bm{\theta}^{\T}} \log f(y^{(h)} \mid \bm{x}^{(h)}; \bm{\theta}^*) \ \bigg| \ \bm{z} \bigg\}
\\
& \phantom{\sum_{h=1}^H \E_{\bm{z}} \bigg[} + \E \bigg\{ \sum_{k=1}^H d^{(k)} t^{(k)} - t^{(h)} w^{(h)}(\bm{z}; \bm{\alpha}^{\dagger}) \ \bigg| \ \bm{z} \bigg\} \bigg\{-\dfrac{\partial^2}{\partial \bm{\theta} \partial \bm{\theta}^{\T}} g^{(h)}(\bm{x}^{(h)},\bm{z};\bm{\theta}^*,\bm{\beta}^{\dagger}) \bigg\} \bigg],
\end{align*}
and if either model is correct, it becomes $\bm{A}(\bm{\theta}^*,\bm{\alpha}^{\dagger})$ on the basis of the definition of \eqref{Adef}. This is because, if the model for the assignment variable is correct, then $\E \{ t^{(h)} w^{(h)}(\bm{z}; \bm{\alpha}^{\dagger}) \mid \bm{z} \} = \E ( \sum_{k=1}^H d^{(k)} t^{(k)} \mid \bm{z} )$, whereas if the model for the outcome variable is correct, then $\E \{ -\partial^2 \log f(y^{(h)}\mid \bm{x}^{(h)}; \bm{\theta}^*) /\partial \bm{\theta}\partial \bm{\theta}^{\T}\mid \bm{z}\} = -\partial^2 g^{( h)}(\bm{x}^{(h)},\bm{z};\bm{\theta}^*,\bm{\beta}^{\dagger}) / \partial \bm{\theta}\partial \bm{\theta}^{\T}$. Thus, letting 
\begin{align}
\bm{C}_1(\bm{\theta},\bm{\alpha},\bm{\beta}) = \bm{A}(\bm{\theta},\bm{\alpha})^{-1} \E \bigg\{ \dfrac{\partial}{\partial \bm{\alpha}^{\T}} \bm{m}(\bm{u}; \bm{\theta}, \bm{\alpha}, \bm{\beta}) \bigg\} \bm{I}_1 (\bm{\alpha})^{-1}
\label{C1def}
\end{align}
and
\begin{align}
\bm{C}_2(\bm{\theta},\bm{\alpha},\bm{\beta}) = \bm{A}(\bm{\theta},\bm{\alpha})^{-1} \E \bigg\{ \dfrac{\partial}{\partial \bm{\beta}^{\T}} \bm{m}(\bm{u}; \bm{\theta}, \bm{\alpha}, \bm{\beta}) \bigg\} \bm{I}_2 (\bm{\beta})^{-1},
\label{C2def}
\end{align}
it can be seen that the inverse of \eqref{mat1} is
\begin{align} 
\begin{pmatrix}
\bm{A}(\bm{\theta}^*,\bm{\alpha}^{\dagger})^{-1} & \bm{C}_1(\bm{\theta}^*,\bm{\alpha}^{\dagger},\bm{\beta}^{\dagger})& \bm{C}_2(\bm{\theta}^*,\bm{\alpha}^{\dagger},\bm{\beta}^{\dagger})
\\
\bm{O} & \bm{I}_1(\bm{\alpha}^{\dagger})^{-1} & \bm{O}
\\
\bm{O} & \bm{O}& \bm{I}_2(\bm{\beta}^{\dagger})^{-1}
\end{pmatrix}.
\label{mat2}
\end{align}
Note that in \eqref{mat1}, the terms in the (1,2) block are
\begin{align*}
& \E \bigg\{ -\dfrac{\partial}{\partial \bm{\alpha}^{\T}} \bm{m}(\bm{u}; \bm{\theta}^*, \bm{\alpha}^{\dagger}, \bm{\beta}^{\dagger}) \bigg\}
\\ 
& = \sum_{h=1}^H \E \bigg[ -\dfrac{\partial}{\partial \bm{\alpha}^{\T}} t^{(h)} w^{(h)}(\bm{z}; \bm{\alpha}^{\dagger}) \bigg\{ \dfrac{\partial}{\partial \bm{\theta}} \log f(y^{(h)} \mid \bm{x}^{(h)}; \bm{\theta}^*) - \dfrac{\partial}{\partial \bm{\theta}} g^{(h)}(\bm{x}^{(h)},\bm{z};\bm{\theta}^*,\bm{\beta}^{\dagger})\bigg\} \bigg]
\\ 
& = \sum_{h=1}^H \E_{\bm{z}} \bigg[ \E \bigg\{ -\dfrac{\partial}{\partial \bm{\alpha}^{\T}} t^{(h)} w^{(h)}(\bm{z}; \bm{\alpha}^{\dagger}) \ \bigg| \ \bm{z} \bigg\} 
\\
& \phantom{= \sum_{h=1}^H \E_{\bm{z}} \bigg(} \E \bigg\{ \dfrac{\partial}{\partial \bm{\theta}} \log f(y^{(h)} \mid \bm{x}^{(h)}; \bm{\theta}^*) - \dfrac{\partial}{\partial \bm{\theta}} g^{(h)}(\bm{x}^{(h)},\bm{z};\bm{\theta}^*,\bm{\beta}^{\dagger}) \ \bigg| \ \bm{z} \bigg\} \bigg]
\\
& = \bm{0}_q,
\end{align*}
if the model for the outcome variable is correct, and the terms in the (1,3) block are
\begin{align*}
& \E \bigg\{ -\dfrac{\partial}{\partial \bm{\beta}^{\T}} \bm{m}(\bm{u}; \bm{\theta}^*, \bm{\alpha}^{\dagger}, \bm{\beta}^{\dagger})^{\T} \bigg\}
\\
& = \sum_{h=1}^H \E \bigg[ \bigg\{ \sum_{k=1}^H d^{(k)} t^{(k)} - t^{(h)} w^{(h)}(\bm{z}; \bm{\alpha}^{\dagger}) \bigg\} \dfrac{\partial^2}{\partial \bm{\beta} \partial \bm{\theta}^{\T}} g^{(h)}(\bm{x}^{(h)},\bm{z};\bm{\theta}^*,\bm{\beta}^{\dagger}) \bigg]
\\
& = \sum_{h=1}^H \E_{\bm{z}} \bigg[ \E \bigg\{ \sum_{k=1}^H d^{(k)} t^{(k)} - t^{(h)} w^{(h)}(\bm{z}; \bm{\alpha}^{\dagger}) \ \bigg| \ \bm{z} \bigg\} \dfrac{\partial^2}{\partial \bm{\beta} \partial \bm{\theta}^{\T}} g^{(h)}(\bm{x}^{(h)},\bm{z};\bm{\theta}^*,\bm{\beta}^{\dagger}) \bigg]
\\
& = \bm{0}_r,
\end{align*}
if the model for the assignment variable is correct. This means that
\begin{align*}
\bm{C}_1(\bm{\theta}^*,\bm{\alpha}^{\dagger},\bm{\beta}^*) = \bm{C}_2(\bm{\theta}^*,\bm{\alpha}^*,\bm{\beta}^{\dagger}) = \bm{O}.
\end{align*}
Using the representation of \eqref{mat2}, we obtain the following lemma from conventional statistical asymptotic theory.

\begin{lemma} \label{DRlem}
Suppose that $\bm{A}(\bm{\theta},\bm{\alpha})$, $\bm{m} (\bm{u};\bm{\theta},\bm{\alpha},\bm{\beta})$, $\bm{C}_1(\bm{\theta},\bm{\alpha},\bm{\beta})$, and $\bm{C}_2(\bm{\theta},\bm{\alpha},\bm{\beta})$ are defined as in \eqref{Adef}, \eqref{mdef}, \eqref{C1def}, and \eqref{C2def}. Then, if either the model for the outcome variable conditional on the confounding variable or the model for the assignment variable conditional on the confounding variable is correct, the error of the doubly robust estimator is expressed as
\begin{align}
\hat{\bm{\theta}}^{\rm DR} - \bm{\theta}^* 
= \frac{1}{N} \sum_{i=1}^N \bigg\{ & \bm{A}(\bm{\theta}^*,\bm{\alpha}^{\dagger})^{-1} \bm{m}(\bm{u}_i; \bm{\theta}^*, \bm{\alpha}^{\dagger}, \bm{\beta}^{\dagger}) + \bm{C}_1(\bm{\theta}^*,\bm{\alpha}^{\dagger},\bm{\beta}^{\dagger}) \sum_{h=1}^H t_i^{(h)} \frac{\partial}{\partial \bm{\alpha}} \log e^{(h)}(\bm{z}_i; \bm{\alpha}^{\dagger})
\notag \\
& + \bm{C}_2(\bm{\theta}^*,\bm{\alpha}^{\dagger},\bm{\beta}^{\dagger}) \sum_{h=1}^H \dfrac{\partial}{\partial \bm{\beta}} t_i^{(h)} \log p^{(h)} (y_i^{(h)} \mid \bm{z}_i; \bm{\beta}^{\dagger}) \bigg\} \{1+\oP(1)\}. 
\label{DRlem1}
\end{align}
Here, at least one of $\bm{\alpha}^{\dagger}$ or $\bm{\beta}^{\dagger}$ will be the true $\bm{\alpha}^*$ or $\bm{\beta}^*$. In particular, when both models are correct, the error is expressed as
\begin{align}
\hat{\bm{\theta}}^{\rm DR} - \bm{\theta}^* = \dfrac{1}{N} \sum_{i=1}^N \bm{A}(\bm{\theta}^*,\bm{\alpha}^*)^{-1} \bm{m}(\bm{u}_i; \bm{\theta}^*, \bm{\alpha}^*, \bm{\beta}^*) \{1+\oP(1)\}.
\label{DRlem2}
\end{align}
\end{lemma}

As an expansion formula for $\hat{\bm{\theta}}^{\rm DR}$, \eqref{DRlem2} is used to show the local asymptotic efficiency, but \eqref{DRlem1} is used here to derive an information criterion while preserving double robustness, which is unexampled to the best of our knowledge. In order to derive the information criterion using this result and \eqref{bias2}, we will evaluate
\begin{align*}
& \E (b^{\rm limit})
\\
& = \dfrac{1}{N} \sum_{i,j=1}^N \sum_{h=1}^H \E \bigg\{ t_i^{(h)} w^{(h)}(\bm{z}_i;\bm{\alpha}^{\dagger}) \dfrac{\partial}{\partial \bm{\theta}^{\T}} \log f(y_i^{(h)} \mid \bm{x}_i^{(h)}; \bm{\theta}^*) \bm{A}(\bm{\theta}^*,\bm{\alpha}^{\dagger})^{-1} \bm{m}(\bm{u}_j; \bm{\theta}^*, \bm{\alpha}^{\dagger}, \bm{\beta}^{\dagger}) \bigg\}
\\
& \phantom{=} \ + \dfrac{1}{N} \sum_{i,j=1}^N \sum_{h,k=1}^H
\\
& \phantom{= +} \ \ \E \bigg\{ t_i^{(h)} w^{(h)}(\bm{z}_i;\bm{\alpha}^{\dagger}) \dfrac{\partial}{\partial \bm{\theta}^{\T}} \log f(y_i^{(h)} \mid \bm{x}_i^{(h)}; \bm{\theta}^*) \bm{C}_1(\bm{\theta}^*,\bm{\alpha}^{\dagger},\bm{\beta}^{\dagger}) \dfrac{\partial}{\partial \bm{\alpha}} t_j^{(k)} \log e^{(k)}(\bm{z}_j; \bm{\alpha}^{\dagger}) \bigg\}
\\
& \phantom{=} \ + \dfrac{1}{N} \sum_{i,j=1}^N \sum_{h,k=1}^H 
\\
& \phantom{= +} \ \ \E \bigg\{ t_i^{(h)} w^{(h)}(\bm{z}_i;\bm{\alpha}^{\dagger}) \dfrac{\partial}{\partial \bm{\theta}^{\T}} \log f(y_i^{(h)} \mid \bm{x}_i^{(h)}; \bm{\theta}^*) \bm{C}_2(\bm{\theta}^*,\bm{\alpha}^{\dagger},\bm{\beta}^{\dagger}) \dfrac{\partial}{\partial \bm{\beta}} t_j^{(k)} \log p^{(k)} (y_j^{(k)} \mid \bm{z}_j; \bm{\beta}^{\dagger}) \bigg\}.
\end{align*}
First, let us evaluate the sum of expectations in the first term with respect to $h$. When $i\neq j$, the expectation is divided into one for $i$ and one for $j$. If either the model for the assignment variable or the model for the outcome variable is correct, the latter $\E\{\bm{m}(\bm{u}_j; \bm{\theta}^*, \bm{\alpha}^{\dagger}, \bm{\beta}^{\dagger})\}$ is $\bm{0}_q$. Therefore, we only have to consider the case of $i=j$; letting
\begin{align}
& \bm{D}_1(\bm{\theta}^*,\bm{\alpha}^{\dagger},\bm{\beta}^{\dagger}) 
\notag \\
& \equiv \sum_{k,h=1}^H \E_{\bm{z}} \bigg[ \E (t^{(h)} \mid \bm{z}) w^{(h)}(\bm{z};\bm{\alpha}^{\dagger}) \dfrac{\partial}{\partial \bm{\theta}} g^{(k)}(\bm{x}^{(k)},\bm{z};\bm{\theta}^*,\bm{\beta}^{\dagger}) \E \bigg\{ \dfrac{\partial}{\partial \bm{\theta}^{\T}} \log f(y^{(h)} \mid \bm{x}^{(h)}; \bm{\theta}^*) \ \bigg| \ \bm{z} \bigg\} \bigg]
\notag \\
& \phantom{\equiv} \ - \sum_{h=1}^H \E_{\bm{z}} \bigg[ \E (t^{(h)} \mid \bm{z}) w^{(h)}(\bm{z};\bm{\alpha}^{\dagger})^2 \dfrac{\partial}{\partial \bm{\theta}} g^{(h)}(\bm{x}^{(h)},\bm{z};\bm{\theta}^*,\bm{\beta}^{\dagger}) \E \bigg\{ \dfrac{\partial}{\partial \bm{\theta}^{\T}} \log f(y^{(h)} \mid \bm{x}^{(h)}; \bm{\theta}^*) \ \bigg| \ \bm{z} \bigg\} \bigg] \bigg),
\label{defD1}
\end{align}
it can be seen that the sum becomes
\begin{align*}
& {\rm tr} \bigg( \bm{A}(\bm{\theta}^*,\bm{\alpha}^{\dagger})^{-1} \sum_{h=1}^H \E_{\bm{z}} \bigg[ \E \bigg\{ t^{(h)} w^{(h)}(\bm{z};\bm{\alpha}^{\dagger}) \bm{m}(\bm{u}; \bm{\theta}^*, \bm{\alpha}^{\dagger}, \bm{\beta}^{\dagger}) \dfrac{\partial}{\partial \bm{\theta}^{\T}} \log f(y^{(h)} \mid \bm{x}^{(h)}; \bm{\theta}^*) \ \bigg| \ \bm{z} \bigg\} \bigg] \bigg) 
\\
& = \mathrm{tr} [ \bm{A}(\bm{\theta}^*,\bm{\alpha}^{\dagger})^{-1} \{ \bm{B}(\bm{\theta}^*,\bm{\alpha}^{\dagger}) + \bm{D}_1(\bm{\theta}^*,\bm{\alpha}^{\dagger},\bm{\beta}^{\dagger}) \} ].
\end{align*}
Next, let us evaluate the sum of the expectations in the second term with respect to $h$. When $i\neq j$, the expectation for $j$ is $\sum_{k=1}^H \E \{ t_j^{(k)}\partial \log e^{(k)}(\bm{z}_j; \bm{\alpha}^{\dagger})/\partial \bm{\alpha}\} = 0$. Therefore, we only have to consider the case of $i=j$; it can be seen that the sum becomes
\begin{align}
& \bm{D}_2(\bm{\theta}^*,\bm{\alpha}^{\dagger},\bm{\beta}^{\dagger})
\notag \\
& \equiv \sum_{h,k=1}^H \E \bigg\{ t^{(h)} w^{(h)}(\bm{z};\bm{\alpha}^{\dagger}) \dfrac{\partial}{\partial \bm{\theta}^{\T}} \log f(y^{(h)} \mid {\bm{x}}^{(h)}; \bm{\theta}^*) \bm{C}_1(\bm{\theta}^*,\bm{\alpha}^{\dagger},\bm{\beta}^{\dagger}) t^{(k)} \dfrac{\partial}{\partial \bm{\alpha}} \log e^{(k)}(\bm{z};\bm{\alpha}^{\dagger}) \bigg\}
\notag \\
&= \sum_{h=1}^H \mathrm{tr} \bigg( \bm{C}_1(\bm{\theta}^*,\bm{\alpha}^{\dagger},\bm{\beta}^{\dagger})
\notag \\
& \phantom{= \sum_{h=1}^H \mathrm{tr} \bigg(} \ \E_{\bm{z}} \bigg[ \E (t^{(h)} \mid \bm{z}) w^{(h)}(\bm{z}; \bm{\alpha}^{\dagger}) \dfrac{\partial}{\partial \bm{\alpha}} \log e^{(h)}({\bm{z}}; \bm{\alpha}^{\dagger}) \E \bigg\{ \dfrac{\partial}{\partial \bm{\theta}^{\T}} \log f(y^{(h)} \mid \bm{x}^{(h)}; \bm{\theta}^*) \ \bigg| \ \bm{z} \bigg\} \bigg] \bigg).
\label{defD2}
\end{align}
Finally, let us evaluate the sum of the expectations in the third term with respect to $h$. When $i\neq j$, the expectation for $j$ is $\sum_{k=1}^H \E [ \partial \{t_j^{(k)} \log p^{(k)} (y_j^{(k)} \mid \bm{z}_j;\bm{\beta}^{\dagger})\} / \partial \bm{\beta} ] = 0$. Therefore, we only have to consider that case of $i=j$, and it can be seen that the sum becomes
\begin{align}
& \bm{D}_3(\bm{\theta}^*,\bm{\alpha}^{\dagger},\bm{\beta}^{\dagger})
\notag \\
& \equiv \sum_{h,k=1}^H \E \bigg\{ t^{(h)} w^{(h)}(\bm{z}; \bm{\alpha}^{\dagger}) \dfrac{\partial}{\partial \bm{\theta}^{\T}} \log f(y^{(h)} \mid {\bm{x}}^{(h)}; \bm{\theta}^*) \bm{C}_2(\bm{\theta}^*,\bm{\alpha}^{\dagger},\bm{\beta}^{\dagger}) t^{(k)} \dfrac{\partial}{\partial \bm{\beta}} \log p^{(k)} (y^{(k)} \mid \bm{z}; \bm{\beta}^{\dagger}) \bigg\}
\notag \\
& = \sum_{h=1}^H \mathrm{tr} \bigg( \bm{C}_2(\bm{\theta}^*,\bm{\alpha}^{\dagger},\bm{\beta}^{\dagger})
\notag \\
& \phantom{= \sum_{h=1}^H \mathrm{tr} \bigg(} \E_{\bm{z}} \bigg[ \E (t^{(h)} \mid \bm{z}) w^{(h)}(\bm{z};\bm{\alpha}^{\dagger}) \dfrac{\partial}{\partial \bm{\beta}} \log p^{(h)} (y^{(h)} \mid \bm{z}; \bm{\beta}^{\dagger}) \E \bigg\{ \dfrac{\partial}{\partial \bm{\theta}^{\T}} \log f(y^{(h)} \mid \bm{x}^{(h)}; \bm{\theta}^*)\ \bigg| \ \bm{z} \bigg\} \bigg] \bigg).
\label{defD3}
\end{align}
From the above, we arrive at the following theorem.

\begin{theorem}
Suppose that $\bm{A}(\bm{\theta},\bm{\alpha})$, $\bm{B}(\bm{\theta},\bm{\alpha})$, $\bm{D}_1(\bm{\theta}^*,\bm{\alpha}^{\dagger},\bm{\beta}^{\dagger})$, $\bm{D}_2(\bm{\theta}^*,\bm{\alpha}^{\dagger},\bm{\beta}^{\dagger})$, and $\bm{D}_3(\bm{\theta}^*,\bm{\alpha}^{\dagger},\bm{\beta}^{\dagger})$ are defined as in \eqref{Adef}, \eqref{Bdef}, \eqref{defD1}, \eqref{defD2} and \eqref{defD3}. Then, if either the model for the outcome variable conditional on the confounding variable or the model for the assignment variable conditional on the confounding variable is correct, the asymptotic bias of the information criterion for the doubly robust estimation is given by
\begin{align*}
\E (b^{\rm limit}) = \mathrm{tr} [ & \bm{A}(\bm{\theta}^*,\bm{\alpha}^{\dagger})^{-1} \{ \bm{B}(\bm{\theta}^*,\bm{\alpha}^{\dagger}) + \bm{D}_1(\bm{\theta}^*,\bm{\alpha}^{\dagger},\bm{\beta}^{\dagger}) \} + \bm{D}_2(\bm{\theta}^*,\bm{\alpha}^{\dagger},\bm{\beta}^{\dagger}) + \bm{D}_3(\bm{\theta}^*,\bm{\alpha}^{\dagger},\bm{\beta}^{\dagger}) ].
\end{align*}
When the former model is correct, $\bm{D}_2(\bm{\theta}^*,\bm{\alpha}^{\dagger},\bm{\beta}^{\dagger})$ becomes $0$; when the latter model is correct, $\bm{D}_3(\bm{\theta}^*,\bm{\alpha}^{\dagger},\bm{\beta}^{\dagger})$ becomes $0$.
\label{th3}
\end{theorem}

Since this asymptotic bias depends on both the true distribution of the outcome variable conditional on the confounding variable and the true distribution of the assignment variable conditional on the confounding variable, one may think that Theorem \ref{th3} is not practical. However, instead, an empirical estimation can provide
\begin{align*}
& \hat{\bm{D}}_1(\hat{\bm{\theta}}^{\rm DR},\hat{\bm{\alpha}},\hat{\bm{\beta}})
\notag \\
& \equiv \dfrac{1}{N} \sum_{i=1}^N \sum_{k,h=1}^H t_i^{(h)} w^{(h)}(\bm{z}_i; \hat{\bm{\alpha}}) \dfrac{\partial}{\partial \bm{\theta}} g^{(k)}(\bm{x}_i^{(k)},\bm{z}_i;\hat{\bm{\theta}}^{\rm DR},\hat{\bm{\beta}}) \dfrac{\partial}{\partial \bm{\theta}^{\T}} \log f(y_i^{(h)} \mid \bm{x}_i^{(h)}; \hat{\bm{\theta}}^{\rm DR})
\notag \\
& \phantom{\equiv} \ - \dfrac{1}{N} \sum_{i=1}^N \sum_{h=1}^H t_i^{(h)} w^{(h)}(\bm{z}_i; \hat{\bm{\alpha}})^2 \dfrac{\partial}{\partial \bm{\theta}} g^{(h)}(\bm{x}_i^{(h)},\bm{z}_i;\hat{\bm{\theta}}^{\rm DR},\hat{\bm{\beta}}) \dfrac{\partial}{\partial \bm{\theta}^{\T}} \log f(y_i^{(h)} \mid \bm{x}_i^{(h)}; \hat{\bm{\theta}}^{\rm DR}),
\end{align*}
\begin{align*}
& \hat{\bm{D}}_2(\hat{\bm{\theta}}^{\rm DR},\hat{\bm{\alpha}},\hat{\bm{\beta}})
\notag \\
& \equiv \dfrac{1}{N} \sum_{i=1}^N \sum_{h=1}^H \mathrm{tr} \bigg\{ \bm{C}_1(\hat{\bm{\theta}}^{\rm DR},\hat{\bm{\alpha}},\hat{\bm{\beta}}) t_i^{(h)} w^{(h)}(\bm{z}_i; \hat{\bm{\alpha}}) \dfrac{\partial}{\partial \bm{\alpha}} \log e^{(h)}(\bm{z}_i; \hat{\bm{\alpha}}) \dfrac{\partial}{\partial \bm{\theta}^{\T}} \log f(y_i^{(h)} \mid \bm{x}_i^{(h)}; \hat{\bm{\theta}}^{\rm DR}) \bigg\},
\end{align*}
\begin{align*}
& \hat{\bm{D}}_3(\hat{\bm{\theta}}^{\rm DR},\hat{\bm{\alpha}},\hat{\bm{\beta}}) 
\notag \\
& \equiv \dfrac{1}{N} \sum_{i=1}^N \sum_{h=1}^H \mathrm{tr} \bigg\{ \bm{C}_2(\hat{\bm{\theta}}^{\rm DR},\hat{\bm{\alpha}},\hat{\bm{\beta}}) t_i^{(h)} w^{(h)}(\bm{z}_i; \hat{\bm{\alpha}}) \dfrac{\partial}{\partial \bm{\beta}} \log p^{(h)} (y_i^{(h)} \mid \bm{z}_i; \hat{\bm{\beta}}) \dfrac{\partial}{\partial \bm{\theta}^{\T}} \log f(y_i^{(h)} \mid \bm{x}_i^{(h)}; \hat{\bm{\theta}}^{\rm DR}) \bigg\}, 
\end{align*}
which are consistent for $\bm{D}_1(\bm{\theta}^*,\bm{\alpha}^{\dagger},\bm{\beta}^{\dagger})$, $\bm{D}_2(\bm{\theta}^*,\bm{\alpha}^{\dagger},\bm{\beta}^{\dagger})$, and $\bm{D}_3(\bm{\theta}^*,\bm{\alpha}^{\dagger},\bm{\beta}^{\dagger})$, respectively. Note that if we had used $\sum_{k=1}^Hd^{(k)}\E(t^{(k)}\mid\bm{z})/\E(t^{(h)}\mid\bm{z})$ instead of $w^{(h)}(\bm{z})$ in \eqref{risk}, we would not have these consistent estimators, so the asymptotic bias in that case is not practical. From the above, we propose
\begin{align}
\mathrm{DRIC} \equiv & -2\sum_{i=1}^N \sum_{h=1}^H \dfrac{t_i^{(h)}}{e^{(h)}(\bm{z}_i; \hat{\bm{\alpha}})} \log f(y_i^{(h)} \mid \bm{x}_i^{(h)}; \hat{\bm{\theta}}^{\rm DR})
\notag \\
& + 2 \mathrm{tr} [ \bm{A}(\hat{\bm{\theta}}^{\rm DR},\hat{\bm{\alpha}})^{-1} \{ \bm{B}(\hat{\bm{\theta}}^{\rm DR},\hat{\bm{\alpha}}) + \hat{\bm{D}}_1(\hat{\bm{\theta}}^{\rm DR},\hat{\bm{\alpha}},\hat{\bm{\beta}}) \} + \hat{\bm{D}}_2(\hat{\bm{\theta}}^{\rm DR},\hat{\bm{\alpha}},\hat{\bm{\beta}}) + \hat{\bm{D}}_3(\hat{\bm{\theta}}^{\rm DR},\hat{\bm{\alpha}},\hat{\bm{\beta}}) ]
\label{DRIC}
\end{align}
as an information criterion for the doubly robust estimation.

\section{Numerical experiment}
\label{sec5}
\subsection{Estimation of various average treatment effects in continuous outcomes}
\label{sec5_1}
Let us consider a model with a treatment group represented by $h=1$ and a control group represented by $h=2$, i.e. $H=2$, where the outcome variable $y^{(h)}\ (\in\mathbb{R})$ is observed for either of them. As the explanatory variables that compose the regression structure, we suppose that $\bm{x}=(x_1,x_2,x_3,x_4)^{\T}$ is distributed according to ${\rm N}(\bm{0}_4,\bm{I}_4)$, and $\bm{x}^{(1)}=(x_1,x_2,0,0)^{\T}$ and $\bm{x}^{(2)}=(0,0,x_3,x_4)^{\T}$, which are different for each group. The assignment variable that becomes $1$ when the sample is in the treatment group is represented by $t^{(1)}$, and the assignment variable that becomes $1$ when the sample is in the control group is represented by $t^{(2)}$. We assume that $y^{(h)}$ and $t^{(h)}$ are correlated and that the confounding variable $\bm{z}=(z_1,z_2)^{\T}$ which can explain the correlation follows a Gaussian distribution ${\rm N}(\bm{0}_2,\bm{I}_2)$ independently of $\bm{x}$.

As the model for $t^{(h)}$, we will use a logit model in which $\bm{z}$ is the explanatory variable. Specifically, letting $\bm{\alpha}=(\alpha_1,\alpha_2)^{\T}$ be the parameter, the model is represented by
\begin{align*}
\P(t^{(1)}=1 \mid \bm{z}; \bm{\alpha}) = \frac{1}{1+\exp(\bm{\alpha}^{\T}\bm{z})}, \qquad t^{(2)}=1-t^{(1)}.
\end{align*}
As the model of $y^{(h)}$, we will use a linear regression model in which $(\bm{x}^{(h)},\bm{z})$ is the explanatory variable as the true structure that generates it. Specifically, letting $\bm{\theta}=(\theta_1,\theta_2,\theta_3,\theta_4)^{\T}$ and $\bm{\beta}=(\beta_1,\beta_2)$ be the parameters, the conditional probability density function of $y^{(h)}$ given $\bm{x}^{(h)}$ and $\bm{z}$ can be expressed as 
\begin{align*}
\frac{1}{\sqrt{2\pi(1-\bm{\beta}^{\T}\bm{\beta})}}\exp\Big\{-\frac{1}{2(1-\bm{\beta}^{\T}\bm{\beta})}(y^{(h)}-\bm{\theta}^{\T}\bm{x}^{(h)}-\bm{\beta}^{\T}\bm{z})^2\Big\}.
\end{align*}
Then, we suppose
\begin{align*}
f(y^{(h)}\mid\bm{x}^{(h)};\bm{\theta}) = \frac{1}{\sqrt{2\pi}}\exp\Big\{-\frac{1}{2}(y^{(h)}-\bm{\theta}^{\T}\bm{x}^{(h)})^2\Big\},
\end{align*}
which actually marginalizes $\bm{z}$, as the probability density function of $y^{(h)}$ in the model. Also, letting $\sigma^{(h)2}$ be the associated variance, we suppose
\begin{align*}
p^{(h)}(y^{(h)}\mid\bm{z};\bm{\beta}) = \frac{1}{\sqrt{2\pi\sigma^{(h)2}}}\exp\Big\{-\frac{1}{2\sigma^{(h)2}}(y^{(h)}-\bm{\beta}^{\T}\bm{z})^2\Big\},
\end{align*}
which marginalizes $\bm{x}^{(h)}$, as the conditional probability density function of $y^{(h)}$ to be used in the doubly robust estimation. That is, the partial derivative of the conditional expectation of $\log f(y^{(h)}\mid\bm{x}^{(h)};\bm{\theta})$ with respect to $\bm{\theta}$ is 
\begin{align*}
\frac{\partial}{\partial\bm{\theta}}g^{(h)}(\bm{x}^{(h)},\bm{z};\bm{\theta},\bm{\beta})
= -\frac{1}{2} \frac{\partial}{\partial\bm{\theta}}\{(\bm{\beta}^{\T}\bm{z})^2+\sigma^{(h)2}-2(\bm{\beta}^{\T}\bm{z})(\bm{\theta}^{\T}\bm{x}^{(h)})+(\bm{\theta}^{\T}\bm{x}^{(h)})^2\}
= (\bm{\beta}^{\T}\bm{z}-\bm{\theta}^{\T}\bm{x}^{(h)})\bm{x}^{(h)}.
\end{align*}

The true values of the parameters used to generate the data are $\bm{\theta}^*=(\theta_1^*,\theta_2^*,\theta_3^*,\theta_4^*)^{\T}=(0.5,0,\theta^*,0)^{\T}$, $\bm{\alpha}^*=(\alpha_1^*,\alpha_2^*)^{\T}=(0.3,\alpha^*)^{\T}$, and $\bm{\beta}^*=(\beta_1^*,\beta_2^*)^{\T}=(0.5,\beta^*)^{\T}$. We use $0.5$ or $0.1$ for $\theta^*$, $0$ or $0.3$ for $\alpha^*$, and $0$ or $0.5$ for $\beta^*$. There are three candidate models that use one or both of $x_1$ and $x_2$ in $\bm{x}^{(1)}$, and three candidate models that use one or both of $x_3$ and $x_4$ in $\bm{x}^{(2)}$, for a total of $3\times 3=9$. Specifically, the following variables are used in the models numbered below.
\begin{center}
{\tabcolsep=3mm
\begin{tabular}{cccccccccc}
 & 1 & 2 & 3 & 4 & 5 & 6 & 7 & 8 & 9 \\
$\bm{x}^{(1)}$ & $x_1,\ x_2$ & $x_1,\ x_2$ & $x_1,\ x_2$ & $x_1$ & $x_1$ & $x_1$ & $x_2$ & $x_2$ & $x_2$ \\
$\bm{x}^{(2)}$ & $x_3,\ x_4$ & $x_3$ & $x_4$ & $x_3,\ x_4$ & $x_3$ & $x_4$ & $x_3,\ x_4$ & $x_3$ & $x_4$ \\
\end{tabular}
}
\end{center}
For all models, we set $\alpha_2=\beta_2=0$. That is, if $\alpha^*\neq 0$, the model for the assignment variable has been misspecified, and if $\beta^*\neq 0$, the model for the outcome variable has been misspecified. The target of the estimation is the treatment effect with $(d^{(1)},d^{(2)})=(1,d^*)$. If $d^*=0$, it is the average treatment effect on the treated (ATT), and if $d^*=1$, it is the average treatment effect of the whole sample (ATE). The sample size is $N=100$ or $N=200$, and the number of repetitions is $3000$.

First, let us examine whether the IPWIC penalty term in \eqref{ipwic2} and the DRIC penalty term in \eqref{DRIC} can approximate the true bias represented by \eqref{bias} in Table \ref{tab1}. In this table, we have used Model 5 in line with the fact that Model 5 is true. That is, in QIC$_w$, the penalty term is $2\times 2=4$ in all cases. The table shows that the approximation of the proposed criterion works reasonably well in all cases. As a matter of fact, when we consider the ATT in two groups, the effect of the inverse probability on the bias is reduced, so that even the penalty term of QIC$_w$ gives a reasonably good approximation. Actually, when $d=0$, the Monte Carlo evaluation (MCE) is close to $4$; here, it cannot be concluded whether the penalty term of QIC$_w$ is superior or inferior to the penalty term of the proposed criterion. On the other hand, if we consider the ATE, the inverse probability effect appears, so the MCE at $d=1$ is larger than $4$ and the proposed penalty term is superior. As the number of groups increases, the effect of the inverse probability becomes more pronounced, and the evaluation of the proposed penalty term becomes more meaningful even when the ATT is considered. Note that this subsection only treats the simple two-group case; the multi-group case is treated in the next subsection.

\begin{table}[t!]
\renewcommand{\baselinestretch}{1.2}\selectfont
\caption{Evaluation of bias in continuous outcome models. The MCE columns list the true values evaluated by the Monte Carlo method, and the AE columns list the asymptotic evaluations.}
\begin{center}
\begin{tabular}{cccccccccccccc}
\hline
 & & \multicolumn{5}{c}{IPWIC} & & \multicolumn{5}{c}{DRIC}
\\
 & & \multicolumn{2}{c}{$N=100$} & & \multicolumn{2}{c}{$N=200$} & & \multicolumn{2}{c}{$N=100$} & & \multicolumn{2}{c}{$N=200$} 
\\
 $(d^*,\theta^*,\alpha^*,\beta^*)$ & & MCE & AE & & MCE & AE & & MCE & AE & & MCE & AE \\ 
\hline
 (0, 0.5, 0, 0) & & ~4.06 & 4.30 & & ~4.38 & 4.29 & & 3.47 & 4.48 & & 3.71 & 4.30 \\
 (1, 0.5, 0, 0) & & ~9.14 & 8.24 & & ~9.62 & 8.28 & & 7.90 & 7.41 & & 8.41 & 7.30 \\
 (0, 0.1, 0, 0) & & ~4.87 & 4.27 & & ~3.67 & 4.28 & & 4.26 & 4.43 & & 2.97 & 4.29 \\
 (1, 0.1, 0, 0) & & ~7.00 & 8.24 & & ~8.88 & 8.23 & & 5.99 & 7.42 & & 7.78 & 7.23 \\
 (0, 0.5, 0.3, 0) & & ~3.99 & 4.15 & & ~4.09 & 4.45 & & 3.36 & 4.64 & & 3.47 & 4.45 \\
 (1, 0.5, 0.3, 0) & & ~8.50 & 8.41 & & ~8.82 & 8.49 & & 7.38 & 7.57 & & 7.65 & 7.46 \\
 (0, 0.1, 0.3, 0) & & ~4.58 & 4.50 & & ~4.13 & 4.45 & & 3.94 & 4.60 & & 3.57 & 4.41 \\
 (1, 0.1, 0.3, 0) & & ~7.87 & 8.54 & & ~7.87 & 8.50 & & 6.65 & 7.64 & & 6.69 & 7.46 \\
 (0, 0.5, 0, 0.5) & & ~4.24 & 4.28 & & ~3.33 & 4.30 & & 3.60 & 4.46 & & 2.77 & 4.30 \\
 (1, 0.5, 0, 0.5) & & 10.17 & 8.21 & & 10.02 & 8.25 & & 9.12 & 7.39 & & 8.80 & 7.25 \\
 (0, 0.1, 0, 0.5) & & ~4.40 & 4.27 & & ~4.17 & 4.28 & & 3.65 & 4.44 & & 3.47 & 4.30 \\
 (1, 0.1, 0, 0.5) & & ~7.72 & 8.31 & & ~8.74 & 8.25 & & 6.60 & 7.45 & & 7.69 & 7.27 \\
\hline
\end{tabular}
\end{center}
\label{tab1}
\end{table}

Tables \ref{tab2_1} and \ref{tab2_2} compare the results of model selection for the settings where the bias evaluation of the proposed criterion works well, with and without model misspecification. Since the propensity score is unknown, we consider IPWIC in \eqref{ipwic2} and DRIC in \eqref{DRIC} to be the proposed criteria and QIC$_w^{\rm IPW}$ using inverse-probability-weighted estimation in QIC$_w$ and QIC$_w^{\rm DR}$ using doubly robust estimation in QIC$_w$ to be the comparison targets. As models, we examine the nine types described above and select the optimal one for each criterion. As the main index to measure the goodness of the criteria, we use the empirical estimator of the risk function in \eqref{risk} minus a constant independent of the model selection, specifically, the average of $3000$ calculations of 
\begin{align}
-2 \sum_{i=1}^N \sum_{h=1}^H \tilde{t}_i^{(h)} w^{(h)}(\tilde{\bm{z}}_i) \log f(\tilde{y}_i^{(h)} \mid \tilde{\bm{x}}_i^{(h)}; \hat{\bm{\theta}}) +2 \sum_{i=1}^N \sum_{h=1}^H \tilde{t}_i^{(h)} w^{(h)}(\tilde{\bm{z}}_i) \log f(\tilde{y}_i^{(h)} \mid \tilde{\bm{x}}_i^{(h)}; \bm{\theta}^*).
\label{erisk}
\end{align}
Comparing the existing criteria with the corresponding proposed criteria, i.e., QIC$_w^{\rm DR}$ and DRIC, and QIC$_w^{\rm IPW}$ and IPWIC, we see that the proposed criteria are superior in all cases, including the case of $d^*=0$ where QIC$_w$ should be somewhat appropriate. According to the selection probabilities, we can see that when $d^*=0.5$ or $d^*=1$, the existing criteria overselect Model 2 or Model 4 and lead to overfitting, and that the effect of underestimating the penalty term is strongly apparent. Note that model misspecification does not degrade performance as the theory suggests and that DRIC is always superior to IPWIC. 

\begin{table}[p]
\renewcommand{\baselinestretch}{1.2}\selectfont
\caption{Comparison of DRIC, IPWIC and QIC$_w$ in continuous outcome models in the absence of model misspecification. The RISK column lists the weighted divergences ($\times 10^2$) determined by \eqref{erisk} and the numbered columns $j\ (\in\{1,2,\ldots,9\})$ evaluate the probability (\%) that the $j$-th model is selected.}
\begin{center}
\begin{tabular}{cccccccccccccc}
\hline
 $(d^*,\theta^*)$ & & RISK & 1 & 2 & 3 & 4 & 5 & 6 & 7 & 8 & 9 \\ 
\hline
 & DRIC &
 1.65 & ~1.9 & 14.7 & 0.0 & ~9.7 & 73.7 & ~0.0 & 0.0 & 0.0 & 0.0 \\
\multirow{2}{*}{(0, 1)} & QIC$_w^{\rm DR}$ &
 1.67 & ~1.9 & 14.6 & 0.0 & 11.1 & 72.4 & ~0.0 & 0.0 & 0.0 & 0.0 \\
 & IPWIC &
 1.77 & ~2.3 & 14.4 & 0.0 & 13.6 & 69.7 & ~0.0 & 0.0 & 0.0 & 0.0 \\
 & QIC$_w^{\rm IPW}$ &
 1.79 & ~2.7 & 13.8 & 0.0 & 15.4 & 68.1 & ~0.0 & 0.0 & 0.0 & 0.0 \\
\hline
 & DRIC &
 2.26 & ~2.6 & 15.4 & 0.0 & 12.7 & 69.3 & ~0.0 & 0.0 & 0.0 & 0.0 \\
\multirow{2}{*}{(0.5, 1)} & QIC$_w^{\rm DR}$ &
 2.39 & ~5.3 & 19.1 & 0.0 & 17.3 & 58.3 & ~0.0 & 0.0 & 0.0 & 0.0 \\
 & IPWIC &
 2.58 & ~2.4 & 14.3 & 0.0 & 13.6 & 69.7 & ~0.0 & 0.0 & 0.0 & 0;.0 \\
 & QIC$_w^{\rm IPW}$ &
 2.83 & ~6.9 & 18.7 & 0.0 & 19.8 & 54.6 & ~0.0 & 0.0 & 0.0 & 0.0 \\
\hline
 & DRIC &
 2.93 & ~2.5 & 14.8 & 0.0 & 14.5 & 68.2 & ~0.0 & 0.0 & 0.0 & 0.0 \\
\multirow{2}{*}{(1, 1)} & QIC$_w^{\rm DR}$ &
 3.26 & ~8.2 & 21.4 & 0.0 & 21.1 & 49.3 & ~0.0 & 0.0 & 0.0 & 0.0 \\
 & IPWIC &
 3.42 & ~2.2 & 13.8 & 0.0 & 13.7 & 70.3 & ~0.0 & 0.0 & 0.0 & 0.0 \\
 & QIC$_w^{\rm IPW}$ &
 3.93 & 10.1 & 22.0 & 0.0 & 22.4 & 45.5 & ~0.0 & 0.0 & 0.0 & 0.0 \\
\hline
 & DRIC &
 1.60 & ~1.9 & 13.9 & 0.0 & ~9.4 & 74.3 & ~0.4 & 0.0 & 0.0 & 0.0 \\
\multirow{2}{*}{(0, 0.5)} & QIC$_w^{\rm DR}$ &
 1.61 & ~2.3 & 13.6 & 0.0 & 10.8 & 73.2 & ~0.2 & 0.0 & 0.0 & 0.0 \\
 & IPWIC &
 1.73 & ~2.3 & 13.4 & 0.0 & 13.1 & 71.1 & ~0.1 & 0.0 & 0.0 & 0.0 \\
 & QIC$_w^{\rm IPW}$ &
 1.77 & ~2.7 & 13.1 & 0.0 & 14.9 & 69.2 & ~0.1 & 0.0 & 0.0 & 0.0 \\
\hline
 & DRIC &
 2.15 & ~2.6 & 14.5 & 0.0 & 12.3 & 70.5 & ~0.1 & 0.0 & 0.0 & 0.0 \\
\multirow{2}{*}{(0.5, 0.5)} & QIC$_w^{\rm DR}$ &
 2.30 & ~5.3 & 18.8 & 0.0 & 16.7 & 59.1 & ~0.1 & 0.0 & 0.0 & 0.0 \\
 & IPWIC &
 2.54 & ~2.4 & 13.3 & 0.0 & 13.4 & 70.8 & ~0.1 & 0.0 & 0.0 & 0.0 \\
 & QIC$_w^{\rm IPW}$ &
 2.79 & ~6.2 & 18.8 & 0.0 & 19.7 & 55.2 & ~0.1 & 0.0 & 0.0 & 0.0 \\
\hline
 & DRIC &
 3.12 & ~3.8 & 13.7 & 0.0 & 14.0 & 68.4 & ~0.0 & 0.0 & 0.1 & 0.0 \\
\multirow{2}{*}{(1, 0.5)} & QIC$_w^{\rm DR}$ &
 3.42 & ~8.9 & 20.1 & 0.0 & 20.6 & 50.3 & ~0.0 & 0.0 & 0.0 & 0.0 \\
 & IPWIC &
 3.63 & ~3.3 & 12.8 & 0.0 & 13.0 & 70.8 & ~0.0 & 0.0 & 0.1 & 0.0 \\
 & QIC$_w^{\rm IPW}$ &
 4.15 & 10.8 & 21.2 & 0.0 & 22.2 & 45.8 & ~0.0 & 0.0 & 0.0 & 0.0 \\
\hline
 & DRIC &
 1.67 & ~0.9 & ~9.3 & 6.7 & ~2.8 & 49.0 & 31.3 & 0.0 & 0.0 & 0.0 \\
\multirow{2}{*}{(0, 0.1)} & QIC$_w^{\rm DR}$ &
 1.71 & ~0.9 & ~9.6 & 6.8 & ~3.2 & 48.5 & 31.0 & 0.0 & 0.0 & 0.0 \\
 & IPWIC &
 1.90 & ~1.1 & ~9.6 & 6.1 & ~4.1 & 49.7 & 29.4 & 0.0 & 0.0 & 0.0 \\
 & QIC$_w^{\rm IPW}$ &
 1.95 & ~1.2 & ~9.7 & 6.3 & ~5.2 & 48.8 & 28.8 & 0.0 & 0.0 & 0.0 \\
\hline
 & DRIC &
 2.35 & ~1.2 & 10.4 & 6.5 & ~4.4 & 48.7 & 28.8 & 0.0 & 0.0 & 0.0 \\
\multirow{2}{*}{(0.5, 0.1)} & QIC$_w^{\rm DR}$ &
 2.48 & ~2.8 & 13.9 & 8.8 & ~7.1 & 42.7 & 24.7 & 0.0 & 0.0 & 0.0 \\
 & IPWIC &
 2.78 & ~1.3 & ~9.3 & 5.8 & ~4.6 & 50.3 & 28.7 & 0.0 & 0.0 & 0.0 \\
 & QIC$_w^{\rm IPW}$ &
 2.98 & ~3.5 & 14.4 & 8.8 & ~8.5 & 41.2 & 23.6 & 0.0 & 0.0 & 0.0 \\
\hline
 & DRIC &
 3.04 & ~1.0 & ~8.8 & 5.6 & ~5.3 & 49.8 & 29.4 & 0.0 & 0.0 & 0.0 \\
\multirow{2}{*}{(1, 0.1)} & QIC$_w^{\rm DR}$ &
 3.26 & ~4.5 & 14.9 & 9.0 & ~9.4 & 39.2 & 22.9 & 0.0 & 0.0 & 0.0 \\
 & IPWIC &
 3.54 & ~0.8 & ~8.9 & 4.8 & ~4.6 & 50.8 & 30.1 & 0.0 & 0.0 & 0.0 \\
 & QIC$_w^{\rm IPW}$ &
 3.88 & ~5.6 & 16.5 & 9.5 & 10.8 & 36.7 & 20.9 & 0.0 & 0.0 & 0.0 \\
\hline
\end{tabular}
\end{center}
\label{tab2_1}
\end{table}

\begin{table}[p]
\renewcommand{\baselinestretch}{1.2}\selectfont
\caption{Comparison of DRIC, IPWIC, and QIC$_w$ in continuous outcome models in the presence of model misspecification. The RISK column lists the weighted divergences ($\times 10^2$) determined by \eqref{erisk} and the numbered columns $j\ (\in\{1,2,\ldots,9\})$ evaluate the probability (\%) that the $j$-th model is selected.}
\begin{center}
\begin{tabular}{cccccccccccccc}
\hline
 $(d^*,\theta^*,\alpha^*,\beta^*)$ & & RISK & 1 & 2 & 3 & 4 & 5 & 6 & 7 & 8 & 9 \\ 
\hline
 & DRIC &
 1.62 & ~2.1 & 14.4 & 0.1 & ~9.2 & 73.9 & ~0.2 & 0.0 & 0.0 & 0.0 \\
\multirow{2}{*}{(0, 0.5, 0.3, 0)} & QIC$_w^{\rm DR}$ &
 1.67 & ~2.8 & 13.4 & 0.0 & 11.7 & 71.8 & ~0.3 & 0.0 & 0.0 & 0.0 \\
 & IPWIC &
 1.73 & ~2.3 & 14.1 & 0.0 & 11.6 & 71.8 & ~0.1 & 0.0 & 0.0 & 0.0 \\
 & QIC$_w^{\rm IPW}$ &
 1.80 & ~3.4 & 12.8 & 0.0 & 14.8 & 68.9 & ~0.0 & 0.0 & 0.0 & 0.0 \\
\hline
 & DRIC &
 2.67 & ~2.4 & 15.0 & 0.0 & 14.2 & 68.3 & ~0.0 & 0.0 & 0.0 & 0.0 \\
\multirow{2}{*}{(1, 0.5, 0.3, 0)} & QIC$_w^{\rm DR}$ &
 2.98 & ~8.3 & 22.1 & 0.0 & 19.8 & 49.8 & ~0.0 & 0.0 & 0.0 & 0.0 \\
 & IPWIC &
 3.20 & ~2.0 & 14.0 & 0.0 & 13.2 & 70.7 & ~0.0 & 0.0 & 0.0 & 0.0 \\
 & QIC$_w^{\rm IPW}$ &
 3.71 & 10.3 & 22.8 & 0.0 & 20.7 & 46.2 & ~0.0 & 0.0 & 0.0 & 0.0 \\
\hline
 & DRIC &
 1.59 & ~2.0 & 14.5 & 0.0 & ~9.9 & 74.5 & ~0.1 & 0.0 & 0.0 & 0.0 \\
\multirow{2}{*}{(0, 0.5, 0, 0.5)} & QIC$_w^{\rm DR}$ &
 1.61 & ~2.2 & 14.1 & 0.0 & 11.1 & 72.5 & ~0.1 & 0.0 & 0.0 & 0.0 \\
 & IPWIC &
 1.71 & ~2.4 & 14.0 & 0.0 & 13.4 & 70.4 & ~0.0 & 0.0 & 0.0 & 0.0 \\
 & QIC$_w^{\rm IPW}$ &
 1.75 & ~2.7 & 13.5 & 0.0 & 15.1 & 68.7 & ~0.0 & 0.0 & 0.0 & 0.0 \\
\hline
 & DRIC &
 2.99 & ~2.6 & 14.4 & 0.0 & 14.8 & 68.1 & ~0.0 & 0.0 & 0.0 & 0.0 \\
\multirow{2}{*}{(1, 0.5, 0, 0.5)} & QIC$_w^{\rm DR}$ &
 3.31 & ~8.3 & 20.6 & 0.0 & 21.5 & 49.6 & ~0.0 & 0.0 & 0.0 & 0.0 \\
 & IPWIC &
 3.48 & ~2.6 & 13.1 & 0.0 & 14.0 & 70.3 & ~0.0 & 0.0 & 0.0 & 0.0 \\
 & QIC$_w^{\rm IPW}$ & 4.00 & 10.1 & 21.9 & 0.0 & 22.4 & 45.6 & ~0.0 & 0.0 & 0.0 & 0.0 \\
\hline
 & DRIC &
 1.60 & ~0.6 & 10.0 & 5.7 & ~2.5 & 48.5 & 32.7 & 0.0 & 0.0 & 0.0 \\
\multirow{2}{*}{(0, 0.1, 0.3, 0)} & QIC$_w^{\rm DR}$ &
 1.62 & ~0.8 & ~9.6 & 5.6 & ~3.1 & 48.1 & 32.8 & 0.0 & 0.0 & 0.0 \\
 & IPWIC &
 1.73 & ~0.6 & 10.0 & 5.5 & ~3.6 & 49.2 & 31.1 & 0.0 & 0.0 & 0.0 \\
 & QIC$_w^{\rm IPW}$ &
 1.79 & ~1.1 & ~9.3 & 5.5 & ~4.9 & 47.7 & 31.5 & 0.0 & 0.0 & 0.0 \\
\hline
 & DRIC &
 3.02 & ~0.9 & 11.3 & 6.0 & ~5.0 & 47.5 & 29.3 & 0.0 & 0.0 & 0.0 \\
\multirow{2}{*}{(1, 0.1, 0.3, 0)} & QIC$_w^{\rm DR}$ &
 3.22 & ~4.2 & 15.5 & 9.2 & 10.3 & 38.1 & 22.7 & 0.0 & 0.0 & 0.0 \\
 & IPWIC &
 3.54 & ~0.6 & 10.4 & 5.5 & ~4.1 & 49.9 & 29.5 & 0.0 & 0.0 & 0.0 \\
 & QIC$_w^{\rm IPW}$ &
 3.89 & ~5.1 & 17.0 & 9.3 & 11.4 & 36.9 & 20.3 & 0.0 & 0.0 & 0.0 \\
\hline
 & DRIC &
 1.74 & ~0.6 & ~9.2 & 6.1 & ~2.5 & 49.1 & 32.6 & 0.0 & 0.0 & 0.0 \\
\multirow{2}{*}{(0, 0.1, 0, 0.5)} & QIC$_w^{\rm DR}$ &
 1.76 & ~0.6 & ~9.2 & 6.1 & ~3.1 & 48.7 & 32.3 & 0.0 & 0.0 & 0.0 \\
 & IPWIC &
 1.95 & ~0.9 & ~8.9 & 5.8 & ~4.0 & 50.2 & 31.2 & 0.0 & 0.0 & 0.0 \\
 & QIC$_w^{\rm IPW}$ &
 1.98 & ~1.2 & ~8.8 & 5.9 & ~5.2 & 48.6 & 30.3 & 0.0 & 0.0 & 0.0 \\
\hline
 & DRIC &
 3.03 & ~1.1 & 10.0 & 5.7 & ~5.1 & 47.8 & 30.4 & 0.0 & 0.0 & 0.0 \\
\multirow{2}{*}{(1, 0.1, 0, 0.5)} & QIC$_w^{\rm DR}$ &
 3.25 & ~4.0 & 15.5 & 9.2 & 10.2 & 38.5 & 22.6 & 0.0 & 0.0 & 0.0 \\
 & IPWIC &
 3.46 & ~1.0 & ~9.3 & 5.1 & ~4.7 & 49.1 & 30.8 & 0.0 & 0.0 & 0.0 \\
 & QIC$_w^{\rm IPW}$ &
 3.80 & ~4.9 & 16.9 & 9.8 & 11.6 & 36.2 & 20.6 & 0.0 & 0.0 & 0.0 \\
\hline
\end{tabular}
\end{center}
\label{tab2_2}
\end{table}

\subsection{Estimation of average treatment effects in discrete outcomes}
\label{sec5_2}
Here, we deal with a setting similar to that in \cite{HosKS06}, and suppose a model with binary outcome variables. Specifically, letting $h\ (\in\{1,2,3,4,5\})$ be the time, the outcome variable $y^{(h)}\ (\in\{0,1\})$ is to be observed somewhere at that time. The assignment variable that becomes $1$ when the observed time is $h$ is represented by $t^{(h)}$. There is a correlation between $y^{(h)}$ and $t^{(h)}$, and the confounding variable $\bm{z}=(z_1,z_2)^{\T}$ that can explain the correlation follows a uniform distribution: ${\rm Unif}(-\sqrt{3},\sqrt{3})$.

As the model for $t^{(h)}$, we use a multinomial logit model in which $\bm{z}$ is the explanatory variable. Specifically, letting $\bm{\alpha}=(\alpha_1,\alpha_2,\alpha_3,\alpha_4,\alpha_5)^{\T}$ be the parameter, the model is represented by 
\begin{align*}
\P(t^{(h)}=1 \mid \bm{z}; \bm{\alpha}) = \frac{\exp(\alpha_h z_1)}{\exp(\alpha_1 z_1)+\exp(\alpha_2 z_1)+\exp(\alpha_3 z_1)+\exp(\alpha_4 z_1)+\exp(\alpha_5 z_2)}
\end{align*}
with $h\in\{1,2,3,4\}$ and $t^{(5)}=1-\sum_{h=1}^4t^{(h)}$. As the model of $y^{(h)}$ used in the doubly robust estimation, we use a logit model with a random effect and $(1,h,h^2,h^3,z_1,z_2)$ as explanatory variables. Specifically, letting $\bm{\theta}=(\theta_1,\theta_2,\theta_3,\theta_4)^{\T}$ and $\bm{\beta}=(\beta_1,\beta_2)^{\T}$ be the parameters, the model is represented by
\begin{align*}
& \P(y^{(h)}=0\mid\bm{z};\bm{\theta},\bm{\beta})
\\
& = \int_{-\infty}^{\infty} \frac{1}{1+\exp(\theta_1+\theta_2h+\theta_3h^2+\theta_4h^3+\bm{\beta}^{\T}\bm{z}+\epsilon)} \frac{1}{\sqrt{2\pi(1-\bm{\beta}^{\T}\bm{\beta})}} \exp\bigg\{-\frac{\epsilon^2}{2(1-\bm{\beta}^{\T}\bm{\beta})}\bigg\} {\rm d}\epsilon.
\end{align*}
Of course, it holds that $\P(y^{(h)}=1\mid\bm{z};\bm{\theta},\bm{\beta})=1-\P(y^{(h)}=0\mid\bm{z};\bm{\theta},\bm{\beta})$. Letting $\epsilon$ be a random effect that follows ${\rm N}(0,1-\bm{\beta}^{\T}\bm{\beta})$ independently of $\bm{z}$, the model of $y^{(h)}$ is a logit model with the regression function as $\theta_1+\theta_2h+\theta_3h^2+\theta_4h^3+\bm{\beta}^{\T}\bm{\beta}+\epsilon$. We suppose
\begin{align*}
\P(y^{(h)}=0\mid\bm{\theta}) = \int_{-\infty}^{\infty} \frac{1}{1+\exp(\theta_1+\theta_2h+\theta_3h^2+\theta_4h^3+\varepsilon)} \frac{1}{\sqrt{2\pi}} \exp\bigg(-\frac{\varepsilon^2}{2}\bigg) {\rm d}\varepsilon,
\end{align*}
which marginalizes $\bm{z}$, as the probability function of $y^{(h)}$ in the model. In generating data according to $\P(y^{(h)}=0\mid\bm{z};\bm{\theta},\bm{\beta})$ above, the true value of the parameter to be used is the one that satisfies $\theta_1^*+\theta_2^*h+\theta_3^*h^2+\theta_4^*h^3=0.5+\theta^*(h-1)$ in the case of a linear polynomial, $\theta_1^*+\theta_2^*h+\theta_3^*h^2+\theta_4^*h^3=0.5+0.2(h-1)+\theta^*(h-1)^2$ in the case of a quadratic polynomial, and in both cases $(\alpha_1^*,\alpha_2^*,\alpha_3^*,\alpha_4^*,\alpha_5^*)=(0.2,-0.15,\allowbreak -0.25,0.4,\alpha^*)$ and $(\beta_1^*,\beta_2^*)=(0.2,\beta^*)$. We will set $0.05$, $0.01$ or $0.005$ for $\theta^*$, and $0$ or $0.2$ for $\alpha^*$ and $\beta^*$. As candidate models, let us consider a 0th-order polynomial (constant) model with $\theta_2=\theta_3=\theta_4=\alpha_5=\beta_2=0$, a 1st-order polynomial (linear) model with $\theta_3=\theta_4=\alpha_5=\beta_2=0$, a 2nd-order polynomial (quadratic) model with $\theta_4=\alpha_5=\beta_2=0$, and a 3rd-order polynomial (cubic) model with $\alpha_5=\beta_2=0$. That is, in all of these models, if $\alpha^*\neq 0$, then the model for the assignment variable has been misspecified, and if $\beta^*\neq 0$, then the model for the outcome variable has been misspecified. The target of the estimation is the average treatment effect of the whole sample with $(d^{(1)},d^{(2)},d^{(3)},d^{(4)})=(1,1,1,1)$. The sample size is $N=100$ or $N=200$, and the number of repetitions is $3000$.

Now let us examine whether the IPWIC penalty term in \eqref{ipwic2} and the DRIC penalty term in \eqref{DRIC} can approximate the true bias represented by \eqref{bias} in Table \ref{tab3}. In this table, when the true model is linear, the assumed model is also linear, and when the true model is quadratic, the assumed model is also quadratic. That is, in QIC$_w$, the penalty term is $2\times 2=4$ when the model is linear, and $2\times 3=6$ when the model is quadratic. The table confirms that the approximation works reasonably well in all cases. At least, they are much closer to the true bias than 4 or 6 in the evaluation of QIC$_w$.

\begin{table}[t]
\renewcommand{\baselinestretch}{1.2}\selectfont
\caption{Bias evaluation in discrete outcome models. The MCE columns are the true values evaluated by the Monte Carlo method, and the AE columns are the asymptotic evaluations.}
\begin{center}
\begin{tabular}{ccccccccccccc}
\hline
 & & & \multicolumn{4}{c}{linear} & & \multicolumn{4}{c}{quadratic}
\\
 & & & \multicolumn{2}{c}{$N=100$} & \multicolumn{2}{c}{$N=200$} & & \multicolumn{2}{c}{$N=100$} & \multicolumn{2}{c}{$N=200$} 
\\
 & $(\theta^*,\alpha^*,\beta^*)$ & & MCE & AE & MCE & AE & & MCE & AE & MCE & AE \\ 
\hline
 & (0.05, 0, 0) & & 25.19 & 22.10 & 25.43 & 21.64 & & 42.74 & 30.10 & 42.69 & 30.71 \\
 & (0.005, 0, 0) & & 27.03 & 22.07 & 21.68 & 21.65 & & 41.35 & 32.67 & 33.23 & 32.44 \\
\multirow{2}{*}{IPWIC} & (0.05, 0.2, 0) & & 27.03 & 22.14 & 26.26 & 21.67 & & 41.27 & 30.21 & 40.50 & 30.72 \\
 & (0.005, 0.2, 0) & & 31.33 & 22.06 & 23.85 & 21.71 & & 41.46 & 32.79 & 40.09 & 32.46 \\
 & (0.05, 0, 0.2) & & 24.64 & 22.15 & 26.36 & 21.69 & & 41.53 & 29.90 & 41.08 & 30.71 \\
 & (0.005, 0, 0.2) & & 25.71 & 22.00 & 25.37 & 21.61 & & 41.39 & 32.69 & 40.21 & 32.41 \\
\hline
 & (0.05, 0, 0) & & 24.90 & 21.76 & 25.00 & 21.41 & & 43.08 & 30.49 & 40.82 & 30.99 \\
 & (0.005, 0, 0) & & 26.85 & 21.77 & 21.40 & 21.44 & & 40.49 & 32.18 & 32.62 & 32.10 \\
\multirow{2}{*}{DRIC} & (0.05, 0.2, 0) & & 26.39 & 21.77 & 25.78 & 21.42 & & 41.59 & 30.74 & 38.53 & 31.21 \\
 & (0.005, 0.2, 0) & & 30.55 & 21.71 & 23.67 & 21.49 & & 40.45 & 32.31 & 39.27 & 32.11 \\
 & (0.05, 0, 0.2) & & 24.48 & 21.85 & 26.06 & 21.44 & & 40.96 & 30.09 & 38.96 & 31.05 \\
 & (0.005, 0, 0.2) & & 25.42 & 21.69 & 25.07 & 21.41 & & 40.32 & 32.18 & 39.21 & 32.05 \\
\hline
\end{tabular}
\end{center}
\label{tab3}
\end{table}

Table \ref{tab4} compares the results of the model selection for the settings where the bias evaluation of the proposed criterion works well. Since the propensity score is unknown, we consider IPWIC in \eqref{ipwic2} and DRIC in \eqref{DRIC} to be the proposed criteria and QIC$_w^{\rm IPW}$ using inverse-probability-weighted estimation in QIC$_w$ and QIC$_w^{\rm DR}$ using doubly robust estimation in QIC$_w$ as the comparison targets. As models, we examine 0th-, 1st-, 2nd- and 3rd-order polynomials in $h$ as described above and select the optimal one for each criterion. As the main index to measure the goodness of the criteria, the evaluated value of risk in \eqref{erisk} is used as in Section \ref{sec5_1}, and also the selection probability is checked as a reference index. Comparing the existing criteria with the corresponding proposed criteria, i.e., QIC$_w^{\rm DR}$ and DRIC, and QIC$_w^{\rm IPW}$ and IPWIC, we find that the proposed criteria are clearly superior in all cases. According to the selection probabilities, the existing criteria are overfitting by selecting too large an order, and the effect of underestimating the penalty term is rather apparent. Especially when $\theta^*$ is small, one might be concerned that the proposed criterion chooses a first-order model that is not true with considerable probability. However, if the contribution of the quadratic term is small, selecting the true model is not necessarily optimal because of the principle of parsimony. Actually, we can confirm that the risk of the proposed criterion is small even if non-true models are selected to some extent.

\begin{table}[p]
\renewcommand{\baselinestretch}{1.2}\selectfont
\caption{Comparison of DRIC, IPWIC and QIC$_w$ in discrete outcome models. The RISK column lists the weighted divergences determined by \eqref{erisk} and the numbered columns $j\ (\in\{0,1,2,3\})$ evaluate the probability (\%) that the $j$-th order polynomial model is selected.}
\begin{center}
\begin{tabular}{cccccccccccccc}
\hline
 & & \multicolumn{5}{c}{$N=100$} & \multicolumn{5}{c}{$N=200$}
\\
 $(\theta^*,\alpha^*,\beta^*)$ & & RISK & 0 & 1 & 2 & 3 & RISK & 0 & 1 & 2 & 3 \\ 
\hline
 & DRIC &
 22.48 & ~0.5 & 45.6 & 31.4 & 22.5 & 14.30 & ~0.0 & 37.2 & 40.3 & 22.5 \\
\multirow{2}{*}{(0.05, 0, 0)} & QIC$_w^{\rm DR}$ &
 26.48 & ~0.0 & 12.8 & 33.2 & 54.0 & 15.40 & ~0.0 & ~9.2 & 34.9 & 55.9 \\
 & IPWIC &
 16.15 & ~0.5 & 47.7 & 36.5 & 15.3 & 11.61 & ~0.0 & 37.7 & 42.7 & 19.6 \\
 & QIC$_w^{\rm IPW}$ &
 17.71 & ~0.0 & 13.2 & 40.2 & 46.6 & 12.61 & ~0.0 & ~8.4 & 34.8 & 56.8 \\
\hline
 & DRIC &
 19.46 & ~0.4 & 44.8 & 32.8 & 30.0 & 13.97 & ~0.0 & 36.3 & 40.8 & 22.9 \\
\multirow{2}{*}{(0.05, 0.2, 0)} & QIC$_w^{\rm DR}$ &
 24.04 & ~0.0 & 12.4 & 34.2 & 53.4 & 15.15 & ~0.0 & ~8.7 & 35.1 & 56.2 \\
 & IPWIC &
 14.53 & ~0.4 & 47.8 & 37.2 & 14.6 & 11.84 & ~0.0 & 36.8 & 42.1 & 21.1 \\
 & QIC$_w^{\rm IPW}$ &
 16.07 & ~0.0 & 12.1 & 41.7 & 46.2 & 12.86 & ~0.0 & ~8.1 & 35.6 & 56.3 \\
\hline
 & DRIC &
 19.79 & ~0.3 & 45.2 & 32.6 & 21.9 & 13.53 & ~0.0 & 37.1 & 41.4 & 21.5 \\
\multirow{2}{*}{(0.05, 0, 0.2)} & QIC$_w^{\rm DR}$ &
 23.68 & ~0.0 & 11.9 & 33.9 & 54.2 & 15.06 & ~0.0 & ~9.0 & 34.4 & 56.6 \\
 & IPWIC &
 13.84 & ~0.3 & 47.7 & 37.5 & 14.5 & 10.89 & ~0.0 & 37.2 & 43.7 & 19.1 \\
 & QIC$_w^{\rm IPW}$ &
 15.55 & ~0.0 & 12.4 & 41.4 & 46.2 & 12.15 & ~0.0 & ~8.3 & 35.1 & 56.6 \\
\hline
 & DRIC &
 11.37 & 25.0 & 46.6 & 15.5 & 12.9 & ~7.69 & ~8.0 & 62.8 & 17.3 & 11.9 \\
\multirow{2}{*}{(0.01, 0, 0)} & QIC$_w^{\rm DR}$ &
 14.33 & ~3.1 & 20.6 & 25.4 & 50.9 & ~9.46 & ~0.5 & 21.6 & 25.9 & 52.0 \\
 & IPWIC &
 11.97 & 25.7 & 47.3 & 15.0 & 12.1 & ~8.15 & ~8.2 & 53.2 & 16.9 & 11.7 \\
 & QIC$_w^{\rm IPW}$ &
 14.12 & ~3.3 & 20.7 & 24.1 & 51.9 & ~9.85 & ~0.4 & 21.2 & 25.4 & 53.0 \\
\hline
 & DRIC &
 12.13 & 24.0 & 47.8 & 15.0 & 13.2 & ~8.51 & ~8.5 & 62.1 & 16.7 & 12.7 \\
\multirow{2}{*}{(0.01, 0.2, 0)} & QIC$_w^{\rm DR}$ &
 14.07 & ~3.0 & 20.5 & 24.1 & 52.4 & 10.04 & ~0.8 & 22.2 & 26.5 & 50.5 \\
 & IPWIC &
 12.51 & 24.3 & 48.4 & 14.5 & 12.8 & ~8.79 & ~8.7 & 62.3 & 16.7 & 12.3 \\
 & QIC$_w^{\rm IPW}$ &
 14.63 & ~3.1 & 20.5 & 24.1 & 52.3 & 10.38 & ~0.8 & 22.4 & 26.0 & 50.8 \\
\hline
 & DRIC &
 12.10 & 25.0 & 46.6 & 15.6 & 12.8 & ~7.97 & ~8.2 & 62.7 & 18.0 & 11.1 \\
\multirow{2}{*}{(0.01, 0, 0.2)} & QIC$_w^{\rm DR}$ &
 14.10 & ~3.2 & 20.6 & 24.6 & 51.6 & ~9.47 & ~0.8 & 21.5 & 26.6 & 51.1 \\
 & IPWIC &
 12.29 & 26.0 & 46.6 & 15.4 & 12.0 & ~8.32 & ~8.4 & 62.9 & 17.5 & 11.2 \\
 & QIC$_w^{\rm IPW}$ &
 14.54 & ~3.4 & 20.1 & 24.3 & 52.2 & ~9.88 & ~0.8 & 21.4 & 25.9 & 51.9 \\
\hline
 & DRIC &
 10.54 & 35.1 & 40.7 & 13.7 & 10.5 & ~8.24 & 18.8 & 56.3 & 14.2 & 10.7 \\
\multirow{2}{*}{(0.005, 0, 0)} & QIC$_w^{\rm DR}$ &
 12.67 & ~5.7 & 21.4 & 23.4 & 49.5 & ~9.67 & ~1.9 & 21.3 & 25.4 & 51.4 \\
 & IPWIC &
 10.86 & 36.1 & 40.4 & 13.7 & ~9.8 & ~8.42 & 18.9 & 56.8 & 13.9 & 10.4 \\
 & QIC$_w^{\rm IPW}$ &
 13.10 & ~5.9 & 21.5 & 23.8 & 48.8 & ~9.93 & ~2.0 & 21.5 & 24.9 & 51.6 \\
\hline
 & DRIC &
 10.96 & 37.0 & 37.2 & 14.1 & 11.7 & ~9.06 & 19.2 & 54.3 & 15.8 & 10.7 \\
\multirow{2}{*}{\hspace{-3mm}(0.005, 0.2, 0)} & QIC$_w^{\rm DR}$ &
 13.02 & ~6.4 & 18.5 & 24.5 & 50.6 & 10.26 & ~2.2 & 22.7 & 25.1 & 50.0 \\
 & IPWIC &
 11.47 & 37.9 & 37.2 & 13.7 & 11.2 & ~9.31 & 19.3 & 54.5 & 15.8 & 10.3 \\
 & QIC$_w^{\rm IPW}$ &
 13.65 & ~6.3 & 18.7 & 24.9 & 50.1 & 10.62 & ~2.1 & 22.3 & 25.2 & 50.4 \\
\hline
 & DRIC &
 10.88 & 37.4 & 38.8 & 14.3 & ~9.5 & ~8.67 & 18.2 & 56.3 & 13.3 & 12.2 \\
\multirow{2}{*}{\hspace{-3mm}(0.005, 0, 0.2)} & QIC$_w^{\rm DR}$ &
 13.37 & ~6.3 & 19.0 & 25.0 & 49.7 & 10.09 & ~1.6 & 22.8 & 24.4 & 51.2 \\
 & IPWIC &
 11.25 & 38.3 & 39.0 & 13.8 & ~8.9 & ~9.04 & 18.6 & 56.7 & 12.9 & 11.8 \\
 & QIC$_w^{\rm IPW}$ &
 13.84 & ~6.5 & 18.7 & 24.6 & 50.2 & 10.63 & ~1.6 & 22.5 & 24.1 & 51.8 \\
\hline
\end{tabular}
\end{center}
\label{tab4}
\end{table}

\section{Real data analysis}
\label{sec6}
The LaLonde dataset is treated in \cite{Lal86} and is included in the R package Matching. The group that took the U.S. job training program in 1976 is denoted as $t=1$ and the group that did not take the program is denoted as $t=0$. The difference in annual income in 1978 after the training for each group is estimated as the average treatment effect on the treated (ATT) or the average treatment effect of the whole sample (ATE). The confounding variables are age (age), years of education (educ), black (black), Hispanic (hisp), married (married), high school graduate or higher (nodegr), income in 1974 (re74), income in 1975 (re75), zero income in 1974 (u74), zero income in 1975 (u75), i.e. $p=10$, and the outcome variable is income in 1978 (re78). The sample size is $N=445$.

The MineThatData dataset is published by \cite{Hil08} and contains 12 attributes, such as purchase amount and district classification code, of 64000 customers. The purpose here is to estimate the causal effect of an email sent to male customers on the probability of the men purchasing a product. Since the sample size of 64000 is too large a computational load, we use a sample of $N=25508$, where the district code (zip\_code) is not urban, and divide the sample into two groups: a group to which the email was delivered to the men ($t=1$) and a group to which it was not sent ($t=0$). The confounding variables are the number of months since the last purchase (recency), the amount purchased in the last year (history), whether the user purchased men's products in the last year (men's), whether the user purchased women's products in the last year (women's), whether the user became a new user in the last 12 months (newbie), i.e. $p=5$, and the outcome variable is whether the user made a purchase within two weeks of receiving the email (conversion).

For these datasets, we use the same model as in Sections \ref{sec5_1} and \ref{sec5_2}, and assume that the regression structure for each latent variable can be written as a linear sum of confounding variables. The confounding variables are then selected by using the step-up procedure using each criterion; the regression coefficients of the selected variables are shown in Table \ref{tab5}. For the LaLonde dataset, we examined both the ATT-targeted DRIC and QIC$_w$ and the ATE-targeted DRIC and QIC$_w$; while the former showed some differences between the two criteria, the latter showed larger differences. For the MineThatData dataset, the proposed criterion selects three variables less than the existing criterion, from which we can confirm that the same trend as in the numerical experiment appears in the real data analysis. Since we do not know the true structure of the real data, it is impossible to judge the superiority or inferiority of the two criteria; however, we can see that the difference between them is quite large.

\begin{table}[t]
\renewcommand{\baselinestretch}{1.2}\selectfont
\caption{Estimates for the real data given by each criterion. Variables with an estimated value of $0$ imply that they were not selected in the model selection. A value of $0$ attached to the criteria means that it is for ATT, and a value of $1$ means that it is for ATE. The exponential notation is used in the figures, and ``E'' is omitted.}
\begin{center}
(a) Lalonde
\medskip\\
\begin{tabular}{rrrrrrrrrrr}
\hline
 & age & educ & black & hisp & marri & nodeg & re74 & re75 & u74 & u75 \\
\hline
DRIC$^0$ & 0\phantom{.3$_{+1}$} & 1.1$_{+2}$ & --1.3$_{+3}$ & 0\phantom{.4$_{+3}$} & 0\phantom{.3$_{+2}$} & 0\phantom{.5$_{+1}$} & 0\phantom{.8$_{-2}$} & 5.4$_{-2}$ & 0\phantom{.9$_{+2}$} & 0\phantom{.8$_{+2}$} \\
QIC$_w^0$ & --0.3$_{+1}$ & 2.1$_{+2}$ & --2.3$_{+3}$ & --1.4$_{+3}$ & 0\phantom{.3$_{+2}$} & 0\phantom{.5$_{+1}$} & 2.8$_{-2}$ & 0\phantom{.4$_{-2}$} & 0\phantom{.9$_{+2}$} & 0\phantom{.8$_{+2}$} \\
\hline
DRIC$^1$ & 3.5$_{+1}$ & 3.3$_{+2}$ & --9.7$_{+3}$ & 0\phantom{.1$_{+3}$} & 0\phantom{.3$_{+2}$} & 0\phantom{.5$_{+1}$} & 0\phantom{.6$_{-2}$} & 0\phantom{.9$_{-2}$} & 0\phantom{.9$_{+2}$} & 0\phantom{.8$_{+2}$} \\
QIC$_w^1$ & 3.7$_{+1}$ & 3.3$_{+2}$ & --9.7$_{+3}$ & --0.1$_{+3}$ & --2.3$_{+2}$ & --6.5$_{+1}$ & 1.6$_{-2}$ & 7.9$_{-2}$ & 6.9$_{+2}$ & --7.8$_{+2}$ \\
\hline
\end{tabular}
\end{center}
\begin{center}
(b) MineThatData
\medskip\\
\begin{tabular}{rrrrrr}
\hline
 & recency & history & men's & women's & newbie \\
\hline
DRIC$^1$ & --5.8$_{-1}$ & 4.4$_{-3}$ & 0\phantom{.0$_{-1}$} & 0\phantom{.4$_{-1}$} & --1.5 \\
QIC$_w^1$ & --7.2$_{-1}$ & --9.3$_{-3}$ & 6.0$_{-1}$ & 1.4$_{-1}$ & --0.5 \\
\hline
\end{tabular}
\end{center}
\label{tab5}
\end{table}

\section{Extension}
\label{sec7}
\subsection{Generalization of divergence}
\label{sec7_1}
While the doubly robust estimation is literally robust against misspecification of the model, it is not robust against outliers because it uses the log-likelihood as the loss function, and one may feel that calling it robust is inappropriate. Therefore, in this subsection, instead of $\log f(y^{(h)} \mid \bm{x}^{(h)}; \bm{\theta})$, we use the loss function $\zeta (y^{(h)} \mid \bm{x}^{(h)}; \bm{\theta})$. Robust loss functions for outliers include those based on $\beta$-divergence treated by \cite{BasHHJ98} and $\gamma$-divergence treated by \cite{FujE08}. Recently, \cite{HarF21} proposed inverse-probability-weighted and doubly robust estimations using such divergences without discussing model selection. Here, we assume that $\bm{\theta}^*$ satisfies
\begin{align}
\sum_{h,k=1}^{H} \E \bigg\{ d^{(k)} t^{(k)} \frac{\partial}{\partial\bm{\theta}} \zeta (y^{(h)} \mid \bm{x}^{(h)}; \bm{\theta}^*) \bigg\} = \bm{0}_p.
\label{IPWEE3}
\end{align}
Let $\eta^{(h)}(\bm{x}^{(h)},\bm{z};\bm{\theta},\bm{\beta})$ be the expectation of $\zeta (y^{(h)}\mid \bm{x}^{(h)}; \bm{\theta})$ taken by $p^{(h)}(y^{(h)}\mid \bm{z}; \bm{\beta})$.

Then, it can be seen that all the content on doubly robust estimation in this paper applies after replacing $\log f$ with $\zeta$ and $g^{(h)}$ with $\eta^{(h)}$. For example, we can get the doubly robust estimating equation by adding \eqref{DRadd} to the left-hand side of \eqref{IPWEE2}. When the model for the propensity score is correct, the expectation of the quantity in the curly brackets in \eqref{DRadd} is $0$, so the estimation equation converges to \eqref{IPWEE3}. Also, when the model for the outcome variable is correct, the expectations of $t^{(h)} w^{(h)}(\bm{z};\bm{\alpha}^{\dagger}) \zeta (y^{(h)}\mid \bm{x}^{(h)}; \bm{\theta}^*)$ and $t^{(h)} w^{(h)}(\bm{z}; \bm{\alpha}^{\dagger}) \eta^{(h)}(\bm{x}^{(h)},\bm{z};\bm{\theta}^*,\bm{\beta}^{\dagger})$ are equal, so asymptotically, these terms cancel out and the estimating equation still converges to \eqref{IPWEE3}. This means that $\hat{\bm{\theta}}^{\rm DR}$ converges in probability to $\bm{\theta}^*$.

The content in Section \ref{sec4} is only valid when either $\bm{\alpha}^{\dagger}$ or $\bm{\beta}^{\dagger}$ is the true value, so one may be concerned that $\bm{\theta}^*$ is not necessarily the true value, but there is no part which is affected by that. Accordingly, Theorem \ref{th3} holds. The true structure appears in the construction of $\hat{\bm{D}}_1$, $\hat{\bm{D}}_2$ and $\hat{\bm{D}}_3$, but again, whether $\bm{\theta}^*$ is the true value or not has no effect. Actually, $\hat{\bm{D}}_1$, $\hat{\bm{D}}_2$ and $\hat{\bm{D}}_3$ are constructible as long as $\E (t^{(h)}\mid \bm{z})$ and $\E \{ {\partial}\log f(y ^{(h)} \mid \bm{x}^{(h)}; \bm{\theta}) / {\partial \bm{\theta}^{\T}} \mid \bm{z} \}$ appear one by one in product form. As a result, we have devised a criterion which replace $\log f$ with $\zeta$ and $g^{(h)}$ with $\eta^{(h)}$ in \eqref{DRIC}.

\subsection{Generalization of weight function}
\label{sec7_2}
Up to this point, the weight function $w^{(h)}(\bm{z};\bm{\alpha})$ for $t^{(h)}$ has been $\sum_{k=1}^H d^{(k)} e^{(k)}(\bm{z};\bm{\alpha}) / e^{(h)}(\bm{z};\allowbreak\bm{\alpha})$ for both the derivation of the estimator and the derivation of the information criterion. In fact, it is possible to construct similar information criteria as in Sections \ref{sec3} and \ref{sec4} even if we change the weight function in each of the derivations; two examples of this will be presented below.

The first example focuses on the true structure for the observed data, as was addressed in \cite{BabKN17}. Specifically, it corresponds to changing only the weight function used in deriving the information criterion to $\sum_{k=1}^H d^{(k)} e^{(k)}(\bm{z};\bm{\alpha})$. In this case, for example, for the inverse-probability-weighted estimation with a known propensity score, Theorem \ref{th1} holds if we remove $1/e^{(h)}(\bm{z};\bm{\alpha})$ from the definition of $\bm{B}(\bm{\theta},\bm{\alpha})$ in \eqref{Bdef}. Furthermore, when $d^{(1)}=d^{(2)}=\cdots=d^{(H)}=1$, which is a setting that considers the average causal effect of the whole sample, it holds that $\bm{A}(\bm{\theta},\bm{\alpha})=\bm{B}(\bm{\theta},\bm{\alpha})$ because $\sum_{k=1}^H d^{(k)} e^{(k)}(\bm{z};\bm{\alpha}) = \sum_{k=1}^H e^{(k)}(\bm{z};\bm{\alpha}) = 1$. Then, the asymptotic bias in Theorem \ref{th1} is calculated to be $-2p$. In other words, although \cite{PlaBCWS13} does not consider this weight function, the information criterion in this case can be written as \eqref{qicw}.

The second example uses covariate balancing (\citealt{ImaR14}), in which the weight function used to derive the estimator is changed. The basic idea of covariate balancing is not to use the maximum likelihood estimator based on the modeling of $e^{(h)}(\bm{z};\bm{\alpha})$ for providing $\hat{\bm{\alpha}}$ but rather to use the information in $\E\{(\partial/\partial\bm{\theta}) \log f(y^{(h)} \mid \bm{x}^{(h)}; \bm{\theta}) \mid \bm{z}\}$ for it. Let us suppose that this conditional expectation is a linear function of $\bm{z}$. Then, for example, referring to \cite{NinPI20}, the estimator is given by solving
\begin{align*}
\sum_{i=1}^N \sum_{h=1}^{H} t_i^{(h)} w^{(h)}(\bm{z}_i;\bm{\alpha}) \bm{z}_i = \bm{0}_q.
\end{align*}
We will write this as $\hat{\bm{\alpha}}^{\rm CB}$ and its limit as $\bm{\alpha}^{\dagger}$. Subsequently, $\hat{\bm{\theta}}^{\rm CB}$ is obtained by solving
\begin{align}
\frac{1}{N} \sum_{i=1}^N \sum_{h=1}^{H} t_i^{(h)} w^{(h)}(\bm{z}_i;\hat{\bm{\alpha}}^{\rm CB}) \frac{\partial}{\partial\bm{\theta}} \log f(y_i^{(h)} \mid \bm{x}_i^{(h)}; \bm{\theta}) = \bm{0}_p.
\label{CBest}
\end{align}
It can be seen that the expectation of the left-hand side of \eqref{CBest} becomes $\bm{0}_p$ if $\E\{(\partial/\partial\bm{\theta}) \log f(y^{(h)} \mid \bm{x}^{(h)}; \bm{\theta}) \mid \bm{z}\}$ is a linear function of $\bm{z}$. This means that $\hat{\bm{\theta}}^{\rm CB}$ converges in probability to $\bm{\theta}^*$. In this setting, let us derive a criterion like IPWIC in Section \ref{sec3_3}, for example, for the selection of $\bm{x}$. It differs from the one in Section \ref{sec3_3} in that it supposes that $\bm{\alpha}^{\dagger}$ is not necessarily the true value. Letting $\bm{\Lambda}(\bm{\theta},\bm{\alpha}) \equiv \sum_{h=1}^H \E [-t^{(h)} \{\partial w^{(h)}(\bm{z}; \bm{\alpha})/\partial \bm{\alpha}\} \{\partial \log f(y^{(h)} \mid \bm{x}^{(h)}; \bm{\theta})/\partial \bm{\theta}^{\T}\}]$ and $\bm{I}(\bm{\alpha}) \equiv \sum_{h=1}^H \E [t^{(h)} \{\partial w^{(h)}(\bm{z};\bm{\alpha})/\partial\bm{\alpha}\} \bm{z}^{\T}]$, the error for $\hat{\bm{\theta}}^{\rm CB}$ can be evaluated similarly to \eqref{IPWest2} as
\begin{align*}
& \hat{\bm{\theta}}^{\rm CB} - \bm{\theta}
\\
& = \bm{A}(\bm{\theta}^*,\bm{\alpha}^{\dagger})^{-1} \frac{1}{N} \sum_{i=1}^N \sum_{h=1}^H t_i^{(h)}w^{(h)}(\bm{z}_i;\bm{\alpha}^{\dagger}) \bigg\{ \frac{\partial}{\partial\bm{\theta}} \log f(y_i^{(h)}\mid\bm{x}_i^{(h)};\bm{\theta}^*) - \bm{\Lambda}(\bm{\theta}^*,\bm{\alpha}^{\dagger})^{\T} \bm{I}(\bm{\alpha}^{\dagger})^{-1} \bm{z}_i \bigg\}.
\end{align*}
Considering the risk function in \eqref{risk}, the asymptotic bias for the information criterion is given by
\begin{align*}
\E (b^{\rm limit}) = & -\frac{2}{N} \sum_{i,j=1}^N \sum_{h,k=1}^H {\rm tr} \bigg(  \bm{A}(\bm{\theta}^*,\bm{\alpha}^{\dagger})^{-1} \E \bigg[ t_i^{(h)}w^{(h)}(\bm{z}_i;\bm{\alpha}^{\dagger}) t_j^{(k)}w^{(k)}(\bm{z}_j;\bm{\alpha}^{\dagger})
\\
& \bigg\{ \frac{\partial}{\partial\bm{\theta}} \log f(y_i^{(h)}\mid\bm{x}_i^{(h)};\bm{\theta}^*) - \bm{\Lambda}(\bm{\theta}^*,\bm{\alpha}^{\dagger})^{\T} \bm{I}(\bm{\alpha}^{\dagger})^{-1} \bm{z}_i \bigg\} \frac{\partial}{\partial\bm{\theta}^{\T}} \log f(y_j^{(k)}\mid\bm{x}_j^{(k)};\bm{\theta}^*) \bigg] \bigg)
\\
= & -2 \sum_{h=1}^H {\rm tr} \bigg( \bm{A}(\bm{\theta}^*,\bm{\alpha}^{\dagger})^{-1} \E \bigg[ t^{(h)}w^{(h)}(\bm{z};\bm{\alpha}^{\dagger})^2
\\
& \bigg\{ \frac{\partial}{\partial\bm{\theta}} \log f(y^{(h)}\mid\bm{x}^{(h)};\bm{\theta}^*) - \bm{\Lambda}(\bm{\theta}^*,\bm{\alpha}^{\dagger})^{\T} \bm{I}(\bm{\alpha}^{\dagger})^{-1} \bm{z} \bigg\} \frac{\partial}{\partial\bm{\theta}^{\T}} \log f(y^{(h)}\mid\bm{x}^{(h)};\bm{\theta}^*) \bigg] \bigg).
\end{align*}
If we substitute consistent estimators for the unknown quantities appearing here, we obtain the following information criterion:
\begin{align*}
& \mathrm{IPWIC}
\\
& \equiv -2 \sum_{i=1}^N \sum_{h=1}^{H} t_i^{(h)} w^{(h)}(\bm{z}_i;\hat{\bm{\alpha}}^{\rm CB}) \frac{\partial}{\partial\bm{\theta}} \log f(y_i^{(h)} \mid \bm{x}_i^{(h)}; \hat{\bm{\theta}}^{\rm CB})
\\
& \ \phantom{\equiv} + \frac{2}{N} \sum_{i=1}^N \sum_{h=1}^H {\rm tr} \bigg[ \bm{A}(\hat{\bm{\theta}}^{\rm CB},\hat{\bm{\alpha}}^{\rm CB})^{-1}
t_i^{(h)}w^{(h)}(\bm{z}_i;\hat{\bm{\alpha}}^{\rm CB})^2
\\
& \ \phantom{\equiv} \bigg\{ \frac{\partial}{\partial\bm{\theta}} \log f(y_i^{(h)}\mid\bm{x}_i^{(h)};\hat{\bm{\theta}}^{\rm CB}) - \bm{\Lambda}(\hat{\bm{\theta}}^{\rm CB},\hat{\bm{\alpha}}^{\rm CB})^{\T} \bm{I}(\hat{\bm{\alpha}}^{\rm CB})^{-1} \bm{z}_i \bigg\} \frac{\partial}{\partial\bm{\theta}^{\T}} \log f(y_i^{(h)}\mid\bm{x}_i^{(h)};\hat{\bm{\theta}}^{\rm CB}) \bigg].
\end{align*}

\section{Conclusion}
\label{sec8}
In this paper, we have developed an information criterion that can be regarded as an asymptotic unbiased estimator of a certain risk function based on the Kullback-Leibler divergence, in order to estimate a general causal effect, which is not necessarily a basic average treatment effect, for a causal inference model, which is not necessarily of the type that can be represented as a linear one. We have also developed an information criterion that is itself doubly robust. The fact that the criterion is doubly robust implies that either the model of the assignment variable or the model of the outcome variable can be misspecified. Thus, the true distributions of the assignment and outcome variables that usually appear in such information criteria cannot be estimated; however, the problem can be avoided by an empirical estimation.

Although propensity score analysis is rapidly being generalized, the methodology for model selection, even in basic settings, have not been well developed. Nevertheless, since even the fundamental development cannot be covered by standard statistical theory, the results presented in this paper are limited to the basic settings. As a relatively easy extension, we have only dealt with a generalization of the divergence to what can be called a triply robust criterion that is robust to outliers and a generalization of the weight functions to handle covariate balancing propensity scores; further customization of the method is a future challenge. In particular, we think that it is important to make confounding and assignment time-dependent and to handle dynamic treatment regimens (see, for example, \citealt{ChaM13} and \citealt{TsiDHL19}). Also, for dynamic treatment regimens, it is necessary to investigate what marginal structures should be modeled and what confounding variables should be included in the direct model for the outcome variables. The topic of estimating causal effects using propensity scores when the confounding variables are high dimensional is a timely topic in econometrics (e.g., \citealt{BelCFH17}, \citealt{CheCDDHNR18}, \citealt{AthIW18}), but even in this setting, it is obvious that, for an efficient estimation, we must select regression models for the explanatory variables and select confounding variables that directly affect the causal effects themselves when they are not so numerous. Extending the information criterion to deal with this problem is also a future challenge.

\section{Acknowledgement}

Yoshiyuki Ninomiya was supported by JSPS Grants-in-Aid for Scientific Research (16K00050) and ISM Cooperative Research Program (2022-ISMCRP-4404).

\bibliography{List}

\end{document}